\documentclass[review,12pt]{elsarticle}

\biboptions{square,comma,sort&compress}

\usepackage{natbib}
\usepackage{amssymb} % The amssymb package provides various useful mathematical symbols
\usepackage{amsthm} % The amsthm package provides extended theorem environments
\usepackage{amsmath}
\usepackage{mathrsfs}
\usepackage{graphicx}
\usepackage{epstopdf}
\usepackage{float}
\usepackage{caption}
\usepackage{soul}
\usepackage{color,soul} %to highlight text
\usepackage{color} %To highlight references in the text
\usepackage{subcaption}
\usepackage{bm}
\usepackage{bbm}
\usepackage{mathrsfs}
\usepackage{soul}
\usepackage[margin=2cm]{geometry}% to reduce margins
\usepackage{booktabs}
\usepackage{chngpage}
\usepackage{enumitem}%enumeration
\usepackage{subcaption}
\usepackage{multirow}
\usepackage{stackengine} %overlap plots
\usepackage{mhchem} %chemical formulas
\usepackage{hyperref} %links
\usepackage{cleveref}
\usepackage{upgreek}
\usepackage{makecell} %for line breaks inside table cells

\journal{International Journal of Hydrogen Energy}

\makeatletter
\def\@author#1{\g@addto@macro\elsauthors{\normalsize%
    \def\baselinestretch{1}%
    \upshape\authorsep#1\unskip\textsuperscript{%
      \ifx\@fnmark\@empty\else\unskip\sep\@fnmark\let\sep=,\fi
      \ifx\@corref\@empty\else\unskip\sep\@corref\let\sep=,\fi
      }%
    \def\authorsep{\unskip,\space}%
    \global\let\@fnmark\@empty
    \global\let\@corref\@empty  %% Added
    \global\let\sep\@empty}%
    \@eadauthor={#1}
}
\makeatother

\makeatletter
\def\thickhline{%
  \noalign{\ifnum0=`}\fi\hrule \@height \thickarrayrulewidth \futurelet
   \reserved@a\@xthickhline}
\def\@xthickhline{\ifx\reserved@a\thickhline
               \vskip\doublerulesep
               \vskip-\thickarrayrulewidth
             \fi
      \ifnum0=`{\fi}}
\makeatother

\newlength{\thickarrayrulewidth}
\begin{document}

\begin{frontmatter}

%\title{Hydrogen embrittlement of 316plus (EN 1.4420) stainless steel at cryogenic temperatures (77 K and 20 K)}

\title{Cryogenic hydrogen embrittlement of 316plus (EN 1.4420) stainless steel at 77 K and 20 K}

%Example of authors list
\author{W. Li\fnref{SO}}
\author{A. Zafra\fnref{OX}}
\author{L. Armendariz\fnref{OX}}
\author{Z. Wang\fnref{CI}}
\author{W. Bailey\fnref{CI}}
\author{E. Martinez-Pañeda\corref{cor1}\fnref{OX}}
\ead{emilio.martinez-paneda@eng.ox.ac.uk}
\author{S.Afshan\corref{cor1}\fnref{SO}}
\ead{s.afshan@soton.ac.uk}

\address[SO]{Faculty of Engineering and Physical Sciences, University of Southampton, Southampton, SO16 7QF, UK}

\address[OX]{Department of Engineering Science, University of Oxford, Oxford OX1 3PJ, UK}

\address[CI]{Institute of Cryogenics, Department of Mechanical Engineering, University of Southampton, Southampton, SO17 1BJ, UK}

\cortext[cor1]{Corresponding authors.}

\begin{abstract}

\noindent This paper presents the first experimental characterisation of combined hydrogen-temperature effects in 316plus (EN 1.4420), a new austenitic stainless steel for liquid hydrogen (LH$_2$) storage. Uniaxial tensile tests were conducted at room temperature (RT), 77 K and 20 K on uncharged and hydrogen-precharged specimens, complemented by fractography and EBSD-based quantification of strain-induced martensite (SIM). 316plus exhibited cryogenic strengthening at 77 K and 20 K by enhanced SIM formation. Hydrogen did not influence strength at RT or 77 K and caused a modest decrease ($\approx$10\%) at 20 K, keeping 316plus at the upper bound of cryogenic strength for 316L. The presence of hydrogen resulted in significant reductions in ductility at all temperatures, being most severe at 77 and 20K ($\approx$40–50\%). Hydrogen suppressed SIM at 20 K, but SIM fraction did not correlate with ductility reduction. Despite the combined effect of temperature and hydrogen, 316plus retained notable ductility (reduction in area $\approx$30\%).

\begin{figure}[H]
     \centering
         \centering
         \includegraphics[width=0.95\textwidth]{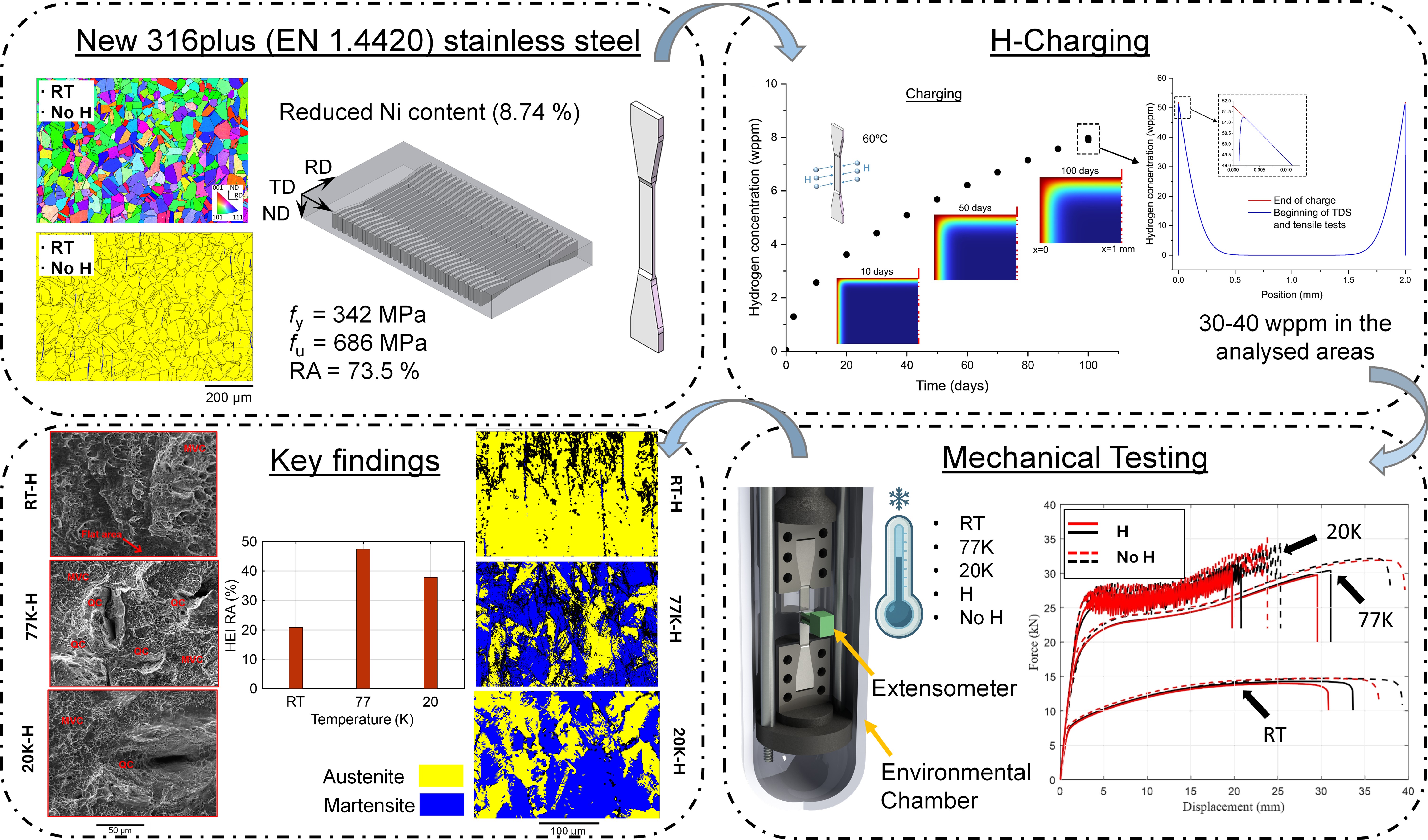}
        \label{fig:GraphicalAbstract}
\end{figure}

\end{abstract}


\end{frontmatter}

%%%%%%%%%%%%%%%%%%%%%%%%%%% INTRODUCTION %%%%%%%%%%%%%%%%%%%%%%%%%%%%%%%%%%%
\section{Introduction}
\label{Introduction}

Hydrogen is increasingly recognised as a key energy carrier for decarbonising hard-to-abate transport sectors, including maritime shipping and aviation, where high energy density, long range and rapid refuelling are essential \cite{suwaileh2025exploring}. In the maritime sector, the International Maritime Organization (IMO) aims to achieve net-zero greenhouse gas (GHG) emissions from international shipping - currently responsible for approximately 3\% of global emissions by 2050, with a commitment to the uptake of zero and near-zero GHG fuels by 2030 \cite{1_imo2023ghgstrategy}. In parallel, liquid hydrogen (LH$_2$) has emerged as a promising fuel for zero-emission aviation, offering the gravimetric energy density required for long-haul flight \cite{cardenas2025physical}. However, deployment across both sectors is limited by gaps in regulatory frameworks and a lack of validated material-performance data. In particular, the storage of hydrogen in cryogenic liquid form (approximately 20–33 K and 1–12.76 bar) poses significant challenges for structural materials, requiring a robust understanding of their behaviour under the combined effects of extremely low temperatures and hydrogen exposure.

Cryogenic LH$_2$ storage requires containment vessels constructed from materials capable of enduring extreme thermal and environmental conditions. Such tanks are typically double-walled, incorporating a vacuum or insulation layer to minimise heat ingress. The inner vessel, which is in direct contact with LH$_2$, is commonly fabricated from austenitic stainless steels \cite{19_wang2021hydrogenstorage,20_afshan2023highperformance}. Alloys such as 304, 304L, 316, 316L and 316LN, are widely used in cryogenic and hydrogen-rich environments due to their excellent corrosion resistance, high fracture toughness and generally favourable ductility at low temperature \cite{IMO2015IGF, IMO2016IGC}. Nevertheless, their mechanical performance under combined hydrogen and cryogenic conditions (below 77 K) remains incompletely understood, owing to the complex interplay between phase stability, microstructural evolution and strain-induced martensitic (SIM) transformation. Material degradation mechanisms - most notably hydrogen embrittlement \cite{chen2025hydrogen} and cryogenic embrittlement \cite{anoop2021review} - represent major concerns for the design and safe operation of LH$_2$ systems.

Hydrogen-assisted degradation of austenitic stainless steels has been examined at room and low temperatures (down to 77 K) \cite{sanmachi2008, michler2008, takaki2016, sanmachi2021}. When hydrogen is introduced in-situ during mechanical testing, these alloys often appear relatively resistant because their low hydrogen diffusivity and stable austenite phase limit the amount of hydrogen that can be transported to deformation sites during loading \cite{Fukuyama2003}. In contrast, pre-charged specimens consistently exhibit a much greater degree of embrittlement \cite{sanmachi2008, michler2008, takaki2016, sanmachi2021}, as hydrogen is already distributed within the microstructure and able to interact directly with dislocations and deformation-induced defects from the onset of loading. These interactions intensify at low temperatures, where reduced stacking-fault energies enhance planar slip and promotes strain-induced $\alpha'$-martensite formation. However, systematic studies on pre-charged austenitic stainless steels under true cryogenic conditions - especially at temperatures as low as 20 K, the boiling point of liquid hydrogen at atmospheric pressure - are currently lacking, leaving the combined hydrogen–temperature response in this regime poorly understood.

To elucidate the temperature-dependent response of austenitic stainless steels and the potential interaction between hydrogen and deformation-induced phase transformations, several studies have examined the behaviour of 316L stainless steel under low-temperature conditions using hydrogen pre-charged specimens. For example, Komatsu et al. \cite{KOMATSU2021} conducted tensile tests on uncharged and hydrogen pre-charged (30-90 wppm) 316L specimens across 293-123 K to assess hydrogen-temperature interactions. Their results showed a continuous increase in ultimate tensile strength with decreasing temperature, attributed to the corresponding rise in SIM fraction, which intensified as temperature decreased and reached its highest levels near the fracture surface before diminishing to zero toward the grips. Hydrogen embrittlement occurred between 233-173 K, manifested by quasi-cleavage fracture features and pronounced strain localisation in the hydrogen-charged specimens. Notably, the SIM fractions were the same in uncharged and hydrogen-charged specimens across all temperatures, indicating that hydrogen did not modify the transformation kinetics and that SIM formation was therefore not the dominant factor governing embrittlement in this regime. Similarly, Álvarez et al. \cite{8_alvarez2023hydrogen} tested 316L at 293 K and 223 K and observed that lowering the temperature increased both yield and ultimate tensile strengths. This strengthening was associated with enhanced martensitic transformation at the lower temperature. Hydrogen pre-charging to 100 wppm led to reduced ductility and more brittle fracture morphologies; however, the SIM evolution showed no dependence on hydrogen. Instead, the amount of SIM was governed primarily by deformation and temperature, indicating that hydrogen did not influence the transformation kinetics. 

Further insight into hydrogen-SIM interactions has been provided by San Marchi et al. \cite{sanmachi2021}, who examined 316L specimens at 293 K and 223 K, with hydrogen contents ranging from 0-140 wppm. At both temperatures, hydrogen increased strength but reduced ductility, with embrittlement severity rising approximately linearly with hydrogen concentration. SIM fractions were substantially higher at 223 K than at 293 K, and hydrogen appeared to promote SIM at early stages of deformation but suppress its growth once the transformation exceeded approximately 20\%. Crucially, the overall amount of SIM did not correlate with ductility loss. The apparent association between SIM and hydrogen-assisted fracture - often inferred by analogy with fully martensitic steels - was therefore argued to be misleading: strain-induced $\alpha'$-martensite is a local transformation product within austenite, whereas fully martensitic steels are bulk microstructures with fundamentally different responses to hydrogen. Hydrogen-assisted degradation in 316L was instead attributed to hydrogen-enhanced planar slip, damage accumulation at slip-band intersections and localised deformation. Bak et al. \cite{Bak2017} compared uncharged and hydrogen pre-charged 316L at equivalent plastic strains and found that hydrogen suppressed overall SIM formation while promoting planar slip and short-range ordering that restricted cross-slip and stabilised the austenite. The resulting laminated $\alpha'$/$\gamma$ structures are consistent with a mechanism in which hydrogen limits the spacial extent and uniformity of SIM transformation.

Building on these observations in the low-temperature regime (approximately 250-125 K), several studies have explored the cryogenic behaviour of 316L in the absence of hydrogen, particularly at temperatures below 77 K \cite{ISHTIAQ2025,Li2023,KIM2024}. These works show that further cooling increases strength and reduces ductility due to more extensive SIM transformation, with the magnitude of transformation strongly dependent on alloy composition, microstructure and deformation mode. At temperatures approaching 4.2 K, martensitic transformation becomes extremely pronounced, producing substantial strengthening \cite{ISHTIAQ2025,Li2023,KIM2024}. However, none of these studies incorporated hydrogen, and to the authors’ knowledge, no systematic evaluation exists of hydrogen-charged 316L (or related grades) at cryogenic temperatures as low as 20 K. The vast majority of the work conducted at 77 K and 20 K has focused exclusively on temperature effects, leaving the combined hydrogen-temperature interaction at cryogenic conditions unexplored. The ductility of 316L alloys can be significantly compromised by the synergistic role of hydrogen and low temperature effects, a phenomenon here termed \emph{cryogenic hydrogen embrittlement}. 

Collectively, previous studies show that although 316-series austenitic stainless steels remain strong candidates for extreme service environments, their mechanical behaviour is highly sensitive to microstructural state and phase stability - both strongly influenced by temperature and hydrogen exposure. Yet a critical knowledge gap persists: the behaviour of these alloys under combined internal hydrogen and cryogenic temperatures approaching 20 K remains largely unexplored, despite its direct relevance to LH$_2$ storage. 

EN 1.4420 (commercially referred to as 316plus) is a recently developed austenitic Cr-Ni-Mo stainless steel characterised by higher chromium and nitrogen contents and relatively lower nickel and molybdenum levels than standard 316 (EN 1.4401) and 316L (EN 1.4404) grades. These compositional modifications are intended to enhance strength, ductility and corrosion resistance. However, the behaviour of this specific grade under cryogenic hydrogen conditions has not yet been systematically investigated. Assessing its response to the combined effects of hydrogen exposure and cryogenic temperatures is therefore essential for evaluating its suitability for LH$_2$ storage applications.

The present work addresses this gap by providing the first systematic assessment of the coupled effects of internal hydrogen and cryogenic temperature on 316plus (EN 1.4420). Uniaxial tensile tests were conducted at room temperature, 77 K and 20 K on both uncharged and hydrogen pre-charged specimens, complemented by SEM fractography to identify operative fracture micromechanisms and EBSD analysis to quantify strain-induced martensite. By integrating these new observations with established findings for 316L, the study clarifies the temperature-dependent interplay between hydrogen, plastic deformation and martensitic transformation in a new alloy specifically designed for LH$_2$ containment.

%%%%%%%%%%%%%%%%%%%%%%%%%%% METHODS %%%%%%%%%%%%%%%%%%%%%%%%%%%%%%%%%%%%%%%%
\section{Materials and experimental methods}
\label{Sec:Experimental}

\subsection{Material and test matrix}
This study was conducted on conventionally manufactured 316plus (EN 1.4420) stainless steel specimens, supplied as hot-rolled plate (1D finish) produced via Electric Arc Furnace and Argon Oxygen Decarburization (E+AOD). The chemical compositions of the plate material, as provided in the mill certificate, is presented in Table \ref{Table_1}. The certified mechanical properties include a 0.2\% proof stress ($f_\mathrm{y}$) of 343 MPa, a 1\% proof stress ($f_\mathrm{1}$) of 390 MPa, an ultimate tensile strength ($f_\mathrm{u}$) of 682 MPa and an elongation to failure ($A50$) of 52\%. 

The $M_{\mathrm{d}}^{30}$ temperature \cite{HATANO2014}, a composition-based metric of the stability of the \(\gamma\)-austenite phase, was determined from the chemical composition of the tested 316plus (EN 1.4420) grade (Table~\ref{Table_1}). In general, a lower $M_{\mathrm{d}}^{30}$ value corresponds to higher \(\gamma\)-phase stability. The calculated $M_{\mathrm{d}}^{30}$ of \(-124\ ^{\circ}\mathrm{C}\) for 316plus is lower than the values typically reported for standard 316L (approximately \(-103\ ^{\circ}\mathrm{C}\) to \(-104\ ^{\circ}\mathrm{C}\) \cite{KOMATSU2021,HATANO2014}), suggesting comparatively enhanced austenite stability in the present steel. Despite the lower nominal Ni content of 8.74 wt.\% compared with approximately 12 wt.\% in conventional 316L \cite{KOMATSU2021,HATANO2014}, the calculated nickel equivalent ($\mathrm{Ni}_{\mathrm{eq}}$) \cite{YAMABE2017} of 316plus is comparable (27.5\%), indicating that the austenite stability is maintained through the balanced alloy composition. Also, the composition-based martensitic start temperature \(M_\mathrm{s}\) \cite{ISHTIAQ2025}, which represents the transformation temperature in the absence of strain, was calculated as \(-285\ ^{\circ}\mathrm{C}\) for the tested specimens. This is lower than typical values reported for standard 316L (around \(-243\ ^{\circ}\mathrm{C}\) to \(-250\ ^{\circ}\mathrm{C}\) \cite{KOMATSU2021, HATANO2014}), further indicating enhanced cryogenic phase stability of the 316plus grade.    

% Table 1 - Chemical composition 
\begin{table}[H]
 \centering
  \caption{Chemical composition of the 316plus (EN 1.4420) material studied.} 
  \label{Table_1}
\begin{tabular}[t]{lcccccccccccc}
\toprule
Element & Cr & Mo & Ni & Mn & Si & Cu & C & N & P & S \\
\midrule
Content (wt.\%) & 20.18 & 0.69 & 8.74 & 1.76 & 0.51 & 0.46 & 0.017 & 0.20 & 0.0333 & 0.001 \\
\bottomrule
\end{tabular}
\end{table}

\begin{figure}[H]
     \centering
         \centering
         \includegraphics[width=1\textwidth]{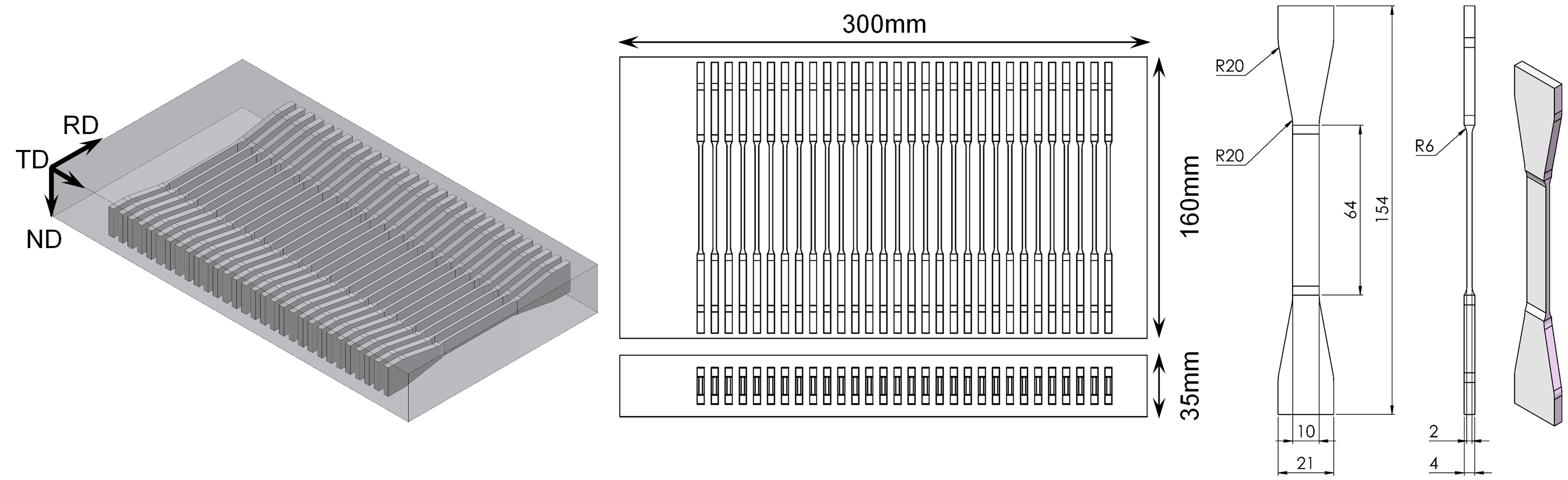}
        \caption{Geometry and extraction of the flat tensile coupon specimens manufactured by electric discharge machining (EDM). Specimens were machined from the 35 mm-thick hot-rolled plate in the rolling direction and had a gauge length of 64 mm with a rectangular cross-section of 10 mm × 2 mm, following BS EN ISO 6892-1 with minor modifications to accommodate the tensile testing grips.}
        \label{fig:Figure_1}
\end{figure}

Tensile coupon specimens were machined from the 35 mm-thick plate in the as-rolled condition using electric discharge machining (EDM). The flat-necked specimens had a gauge length of 64 mm and a rectangular gauge cross-section of 10 mm × 2 mm, as shown in Fig. \ref{fig:Figure_1}. The specimen geometry followed BS EN ISO 6892-1 \cite{4_bseniso6892-1}, with minor modifications to accommodate the tensile testing grips. Tests were performed at room temperature (295 K) and at cryogenic temperatures of 77 K and 20 K, under both uncharged and hydrogen pre-charged conditions.

%Figure 1 - Coupon extraction

Electron Backscatter Diffraction (EBSD) analysis was conducted to characterise the microstructure of the as-received austenitic 316plus (EN 1.4420) plate. The analysis was carried out using a JEOL JSM-7200F Scanning Electron Microscope (SEM), equipped with an Oxford Instruments C-Nano EBSD detector, operated at 20 kV. Sample preparation involved sequential grinding with SiC paper, followed by polishing with 3 $\upmu$m and 1 $\upmu$m water-based diamond suspensions, and a final polish using 0.06 $\upmu$m colloidal silica. During EBSD acquisition, the specimen was tilted at 70°, and a step size of 0.75 $\upmu$m was used. Data were processed using \textit{Aztec Crystal} software. 

\subsection{Hydrogen charging}
\label{Sec:HydrogenCharging}
The tensile specimens were ground to a \#1200 grit surface finish and subsequently electrochemically pre-charged using a customised three-electrode cell containing 5 L of deaerated 3.5 wt.\% NaCl solution maintained at 60 $^\circ$C ($\pm$2 $^\circ$C). A constant cathodic current density of 5 mA/cm$^2$ was applied using two platinum counter electrodes (6 mm diameter rods) and a TTi PL303 power supply, resulting in an average electrochemical potential of -1.3 V, measured against a silver/silver chloride (Ag/AgCl) reference electrode using a Gamry 1010B potentiostat. Due to the low hydrogen diffusivity in 316L stainless steels at 60 $^\circ$C ($\sim$1.6$\times10^{-15}$ m$^2$/s) \cite{BRASS2006}, the samples were charged for 100 days. Throughout the charging period, the pH of the solution was continuously monitored, and the electrolyte was periodically refreshed to maintain stable conditions and minimise sample-to-sample variations in hydrogen uptake. These charging conditions (5 mA/cm$^2$, 60 $^\circ$C, 100 days), previously employed to introduce high hydrogen concentrations in electrochemically pre-charged FCC alloys \cite{QUAN2026,SantosMaldonado2024}, represent an accelerated electrochemical pre-charging procedure rather than a direct simulation of LH$_2$ service exposure. They were selected to introduce a sufficient amount of hydrogen into the 2 mm-thick austenitic stainless-steel specimens, enabling the coupled influence of hydrogen and cryogenic temperature on the mechanical response to be examined. Upon completion of charging, the tensile specimens were immediately immersed in liquid nitrogen to prevent hydrogen loss and stored under cryogenic conditions until mechanical testing.

Thermal desorption spectroscopy (TDS) measurements were performed on 10 mm $\times$ 10 mm $\times$ 2 mm specimens extracted from the gauge section of a non-tested tensile specimen pre-charged under the conditions described above to quantify the hydrogen content introduced during charging. The analysis was conducted using an ultra-high vacuum thermal desorption spectroscopy (UHV-TDS) system equipped with a regularly calibrated Hiden Analytical RC PIC quadrupole mass spectrometer. During hydrogen desorption, the background pressure in the measurement chamber was maintained below ${<}1\!\times10^{-8}$ mbar, providing a sensitivity of $\pm$0.01 wppm. Three independent measurements were carried out using a constant heating rate of 30 $^\circ$C/min over the temperature range 25-850 $^\circ$C, while the hydrogen desorption rate (wppm/s) was continuously recorded as a function of time. The total hydrogen content was determined by integrating the desorption spectra after background subtraction. To minimise hydrogen loss, a dwell time of 30 minutes was maintained between the completion of charging and the start of the TDS analysis. Further details of the UHV-TDS system and measurement protocol are provided in \cite{ZAFRA2023_TDS}. To estimate the hydrogen concentration profile within the specimens at the onset of the TDS and tensile experiments, a 2-D diffusion simulation was performed using COMSOL Multiphysics.

\subsection{Mechanical testing}
\label{Subsec:Mechanical testing}

\begin{figure}[H]
     \centering
         \centering
         \includegraphics[width=1\textwidth]{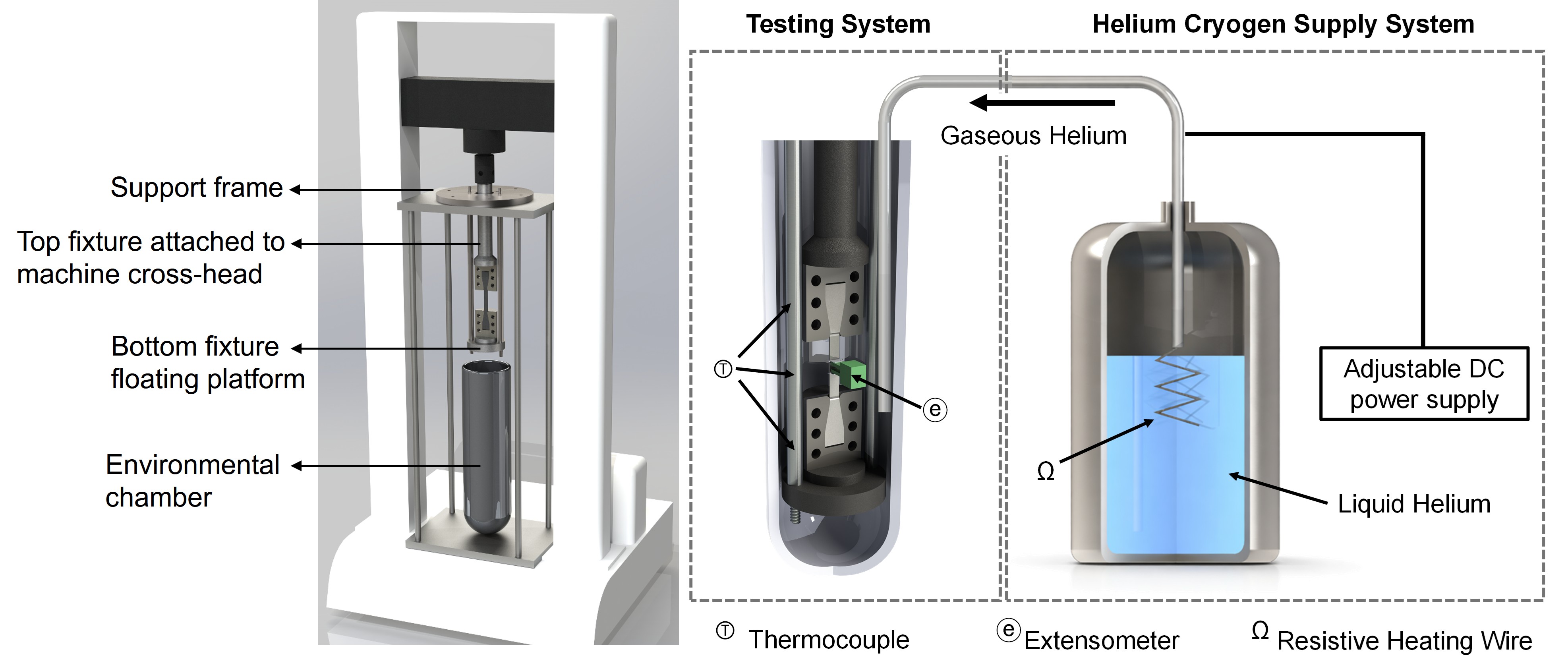}
        \caption{Cryogenic tensile testing setup used for mechanical testing at room temperature, 77 K and 20 K. The custom-built rig consists of a support frame, two end fixtures connected to the testing machine, and specially designed grips to accommodate the flat tensile specimens. Cryogenic testing was performed inside a vacuum-insulated Dewar chamber, where nitrogen or helium gas - generated by boiling liquid nitrogen (LN$_2$) or liquid helium (LHe) - was injected at the base of the chamber to cool the specimen to 77 K or 20 K, respectively. Temperature near the specimen was continuously monitored using a calibrated cryogenic thermometer to ensure stable testing conditions.}
        \label{fig:Figure_4}
\end{figure}

Tensile tests were performed on both uncharged and hydrogen pre-charged specimens at room temperature and at cryogenic temperatures of 77 K, and 20 K. All experiments were carried out in accordance with ISO 6892-1 \cite{4_bseniso6892-1} and ISO 6892-4 \cite{7_bseniso6892-4} standards. Uniaxial tensile testing was conducted using an Instron 50 kN 3369 electromechanical testing machine. A dedicated cryogenic test rig was designed and manufactured to accommodate low-temperature testing. Fig. \ref{fig:Figure_4} shows the overall experimental setup and instrumentation. The rig consisted of a support frame, two end fixtures (one attached to the machine crosshead and one acting as a floating platform at the base), two grips and corresponding cover plates. The custom-built grips featured open slots shaped to match the specimen ends, allowing secure positioning and bolting into the end fixtures using the cover plates. 

Tensile tests at 77 K and 20 K were conducted within an environmental chamber constructed from silvered, vacuum-insulated glass (Dewar type). The chamber was positioned beneath the floating platform of the test rig and fully enclosed the specimen. Cryogenic nitrogen and helium gases were injected into the chamber to cool the specimens to the target temperatures of 77 K and 20 K, respectively. The gases were produced by boiling liquid nitrogen (LN2) and liquid helium (LHe) from storage vessels using an auxiliary heater and were injected at the base of the chamber to ensure efficient heat exchange with the specimen (Fig. \ref{fig:Figure_4}). The temperature near the specimen was continuously monitored using a calibrated cryogenic thermometer to ensure temperature stability throughout testing. Prior to testing, hydrogen pre-charged specimens were removed from liquid nitrogen storage and equilibrated at room temperature for 30 minutes before assembly into the test rig for both room-temperature and cryogenic testing. All tests were performed under displacement control, with the machine crosshead moving at a constant rate of 1 mm/min, corresponding to an initial strain rate of 2.6$\times10^{-4}$ s$^{-1}$. A 10 mm gauge length clip-on extensometer was used to measure local strain, and the applied load was recorded using a calibrated 50 kN load cell.

%Figure 4 - Testing setup

The reduction in area (RA) at the fracture surface was used to directly quantify the true plastic strain at fracture and to compare specimen ductility under different test conditions. Profile images of the fracture surfaces were obtained using a calibrated Olympus BX41M-LED optical microscope. The initial cross-sectional area ($A_\mathrm{0}$) and the area at fracture ($A_\mathrm{f}$) were measured and used to calculate the RA as: RA(\%) = 100×($A_\mathrm{0}$ – $A_\mathrm{f}$)/$A_\mathrm{0}$. A temperature-dependent hydrogen embrittlement index (HEI) was defined to evaluate the effect of hydrogen on ductility: HEI(\%) = 100×($\mathrm{RA_{NoH}}$ – $\mathrm{RA_H}$)/$\mathrm{RA_{NoH}}$, where $\mathrm{RA_{NoH}}$ and $\mathrm{RA_H}$ correspond to the reductions in areas of the uncharged and hydrogen pre-charged specimens, respectively, at a given temperature. Similarly, a cryogenic embrittlement index (CEI) was defined to assess the effect of low temperature on ductility: CEI(\%) = 100×($\mathrm{RA_{RT}}$ – $\mathrm{RA_C}$)/$\mathrm{RA_{RT}}$, where $\mathrm{RA_{RT}}$ and $\mathrm{RA_C}$ correspond to the reduction in area at room and cryogenic temperatures, respectively. Finally, the combined effect of cryogenic temperature and hydrogen on ductility (RA) is measured by defining a \emph{cryogenic hydrogen embrittlement index} (CHEI) defined as CHIE(\%)= 100×($\mathrm{RA_{RT,NoH}}$ – $\mathrm{RA_{C,H}}$)/$\mathrm{RA_{RT,NoH}}$, where $\mathrm{RA_{RT,NoH}}$ and $\mathrm{RA_{C,H}}$ correspond to the reductions in area measured at room temperature without hydrogen and at cryogenic temperature with hydrogen pre-charging, respectively. Higher values of HEI, CEI or CHEI indicate increased susceptibility to embrittlement arising from hydrogen, cryogenic temperature, or their combined effect. This RA-based approach is particularly suitable for ductile metals that experience extensive post-uniform deformation, as opposed to using plastic strain at fracture \cite{8_alvarez2023hydrogen}.

\subsection{Fracture surface and EBSD analysis}
\label{Subsec:Microstructure analysis}

Fractographic analysis was conducted to assess the influence of cryogenic temperature and hydrogen on the fracture mechanism of the tested 316plus specimens. Scanning Electron Microscopy (SEM) images of the fracture surfaces were obtained using a Leo 1450VP SEM equipped with the \textit{Aztec Energy} software. To quantify the local content of strain-induced $\alpha'$-martensite, EBSD analysis was performed on samples extracted from the gauge length - specifically, from the vicinity of the fracture ($\sim$0.5 mm from the fracture edge) and from the uniformly deformed region ($\sim$15 mm away from the fracture edge). Sample preparation involved sequential mechanical grinding with SiC papers, followed by polishing with 3 $\upmu$m and 1 $\upmu$m water-based diamond suspensions, and final polishing with 0.06 $\upmu$m colloidal silica. EBSD scans were acquired using a JEOL JSM-7200F SEM equipped with an Oxford Instruments C-Nano detector, operated at 20 kV with a step size of 1.5 $\upmu$m. Data were collected and processed using \textit{Aztec Crystal} software.

%%%%%%%%%%%%%%%%%%%%%%%%%% RESULTS %%%%%%%%%%%%%%%%%%%%%%%%%%%%%%%
\section{Results}
\label{Sec:Results}

\subsection{Microstructure}

\begin{figure}[H]
     \centering
         \centering
         \includegraphics[width=1\textwidth]{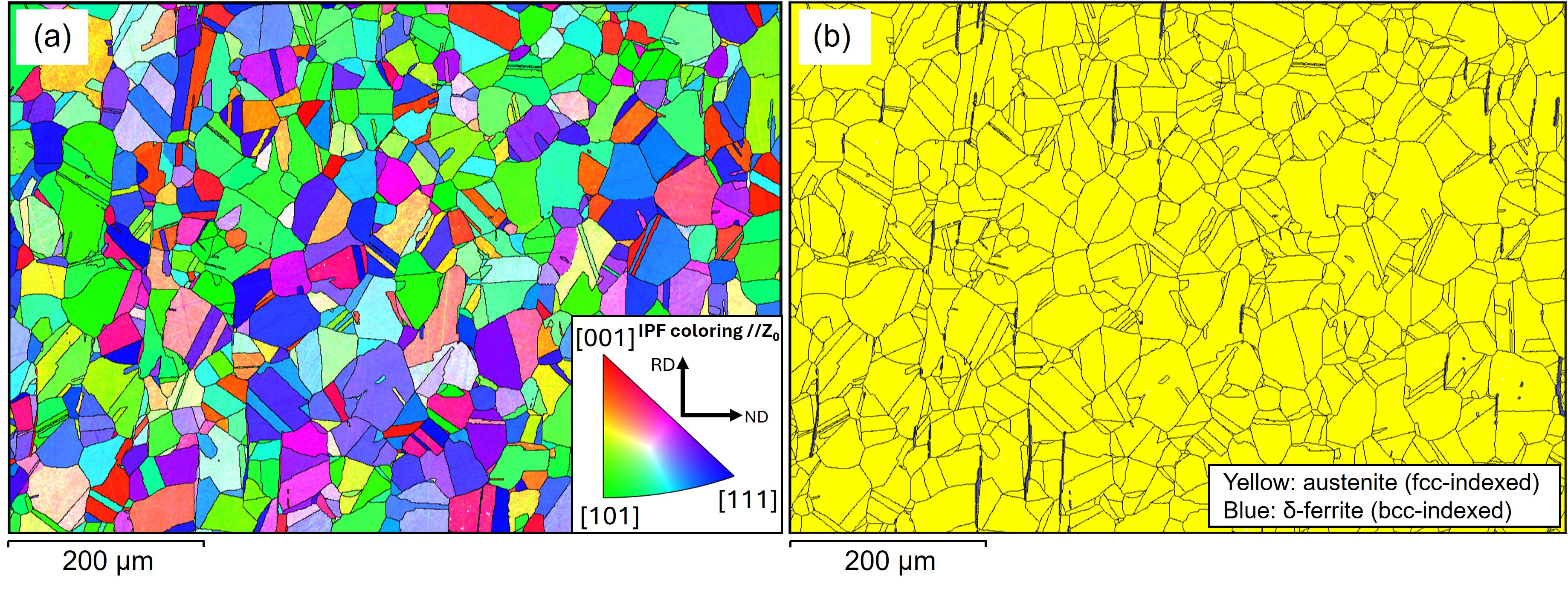}
        \caption{(a) Inverse pole figure (IPF) map and (b) phase map of the as-received 316plus (EN 1.4420) material obtained by EBSD. The microstructure consists predominantly of $\gamma$-austenite with an average grain diameter of $\sim$19 $\upmu$m and a small fraction of $\delta$ ferrite stringers, mainly located along grain boundaries.}
        \label{fig:Figure_2}
\end{figure}

The inverse pole figure (IPF) map and phase map of the as-received 316plus (EN 1.4420) are shown in Fig. \ref{fig:Figure_2}. The grain microstructure in Fig. \ref{fig:Figure_2}(a) reveals an austenitic matrix composed of equiaxed to slightly elongated grains with well-defined boundaries and relatively uniform intragranular orientation. The mean grain diameter is approximately 19 $\upmu$m. Discrete colour bands within some grains indicate the presence of annealing twins, likely formed during hot rolling. The corresponding phase map in Fig. \ref{fig:Figure_2}(b) confirms that more than 99.4\% of the scanned area consists of face-centred cubic (fcc) phase, consistent with $\gamma$-austenite. A small fraction ($\approx$ 0.5\%) of discontinuous body-centred cubic (bcc) phase - likely corresponding to ferrite stringers ($\delta$ ferrite) \cite{8_alvarez2023hydrogen} - was also detected, mainly along grain boundaries.
  
%Figure 2 - 316L+ microstructure

\subsection{Hydrogen concentration profile}

An average hydrogen concentration of 7.9$\pm$1.4 wppm was measured in the 2 mm-thick specimens after 100 days of electrochemical charging using TDS. However, this value represents the average hydrogen content across the entire specimen thickness, calculated with respect to the total mass of the sample. Because the specimens were not charged to saturation, this value does not represent the local hydrogen concentration within the hydrogen-enriched regions of the material.

\begin{figure}[H]
     \centering
         \centering
         \includegraphics[width=1\textwidth]{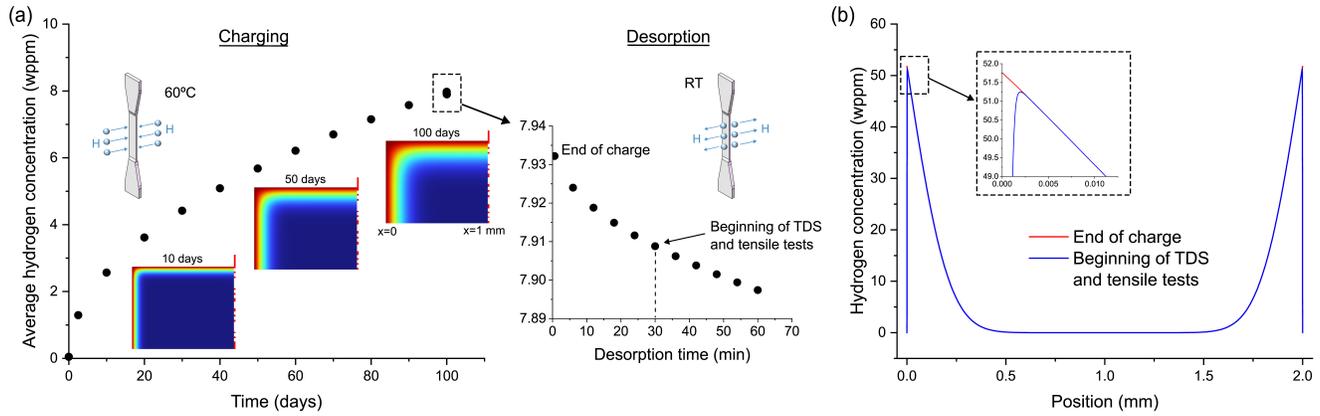}
        \caption{2-D diffusion finite element simulations performed using COMSOL showing (a) the evolution of the average hydrogen concentration in the specimens during charging and desorption, and (b) hydrogen concentration profiles across the specimen thickness at the end of charging and at the start of the TDS and tensile tests (after a 30 min dwell time). Although the experimentally measured average hydrogen concentration was 7.9 wppm, substantially higher hydrogen levels (up to 52 wppm) are present within the hydrogen-enriched near-surface regions examined in the fractographic analyses.}
        \label{fig:Figure_3}
\end{figure}

Therefore, a 2-D diffusion simulation was performed using COMSOL Multiphysics to estimate the hydrogen concentration profile in the specimens at the onset of the TDS and tensile experiments (i.e., 30 min after charging) \cite{diaz2025comsol}. As shown in Fig. \ref{fig:Figure_3}(a), both the charging and desorption stages were modelled. During the charging phase (100 days at 60 $^\circ$C and 5 mA/cm$^2$), the specimen surfaces were assigned a constant hydrogen concentration boundary condition of 52 wppm, which produced an average hydrogen concentration of 7.9 wppm at the end of charging, in excellent agreement with the experimentally measured value. This concentration of 52 wppm, which represents the effective equilibrium solubility under the imposed electrochemical charging conditions, is also consistent with reported hydrogen solubility data for 316L stainless steel (50–140 wppm) \cite{8_alvarez2023hydrogen, sanmachi2021, SanMarchi2012}.

%Figure 3 - Simulations

The subsequent desorption phase (30 min dwell at room temperature) was simulated by using the hydrogen concentration profile at the end of charging as the initial condition and imposing a zero-concentration boundary condition at all external surfaces, representing immediate exposure to a hydrogen-free environment. Temperature-dependent hydrogen diffusivities for 316L stainless steel at 60 $^\circ$C and room temperature were used for the charging and desorption stages, respectively \cite{BRASS2006}, as 316L provides the closest available analogue to 316plus. Because the diffusivity at room temperature is extremely low ($\sim$1.2$\times10^{-16}$ m$^2$/s), only negligible hydrogen escaped during the dwell period preceding TDS and mechanical testing. 

Fig. \ref{fig:Figure_3}(b) shows the hydrogen concentration profiles immediately after charging and after the 30-minute room temperature dwell. The profiles are essentially identical, except for a very thin surface layer (depth $<$2 $\upmu$m) where minor desorption occurred. After 100 days of charging, hydrogen had penetrated to a depth of approximately 0.5 mm from each surface, reaching concentrations of approximately 52 wppm near the surface. Consequently, although the experimentally measured hydrogen concentration was 7.9 wppm, substantially higher hydrogen levels were present within the hydrogen-enriched near-surface regions that were later examined in the fractographic analyses.

\subsection{Mechanical behaviour}
\label{Subsec:Stress-strain curves and fracture surfaces}

\begin{figure}[H]
     \centering
         \centering
         \includegraphics[width=1\textwidth]{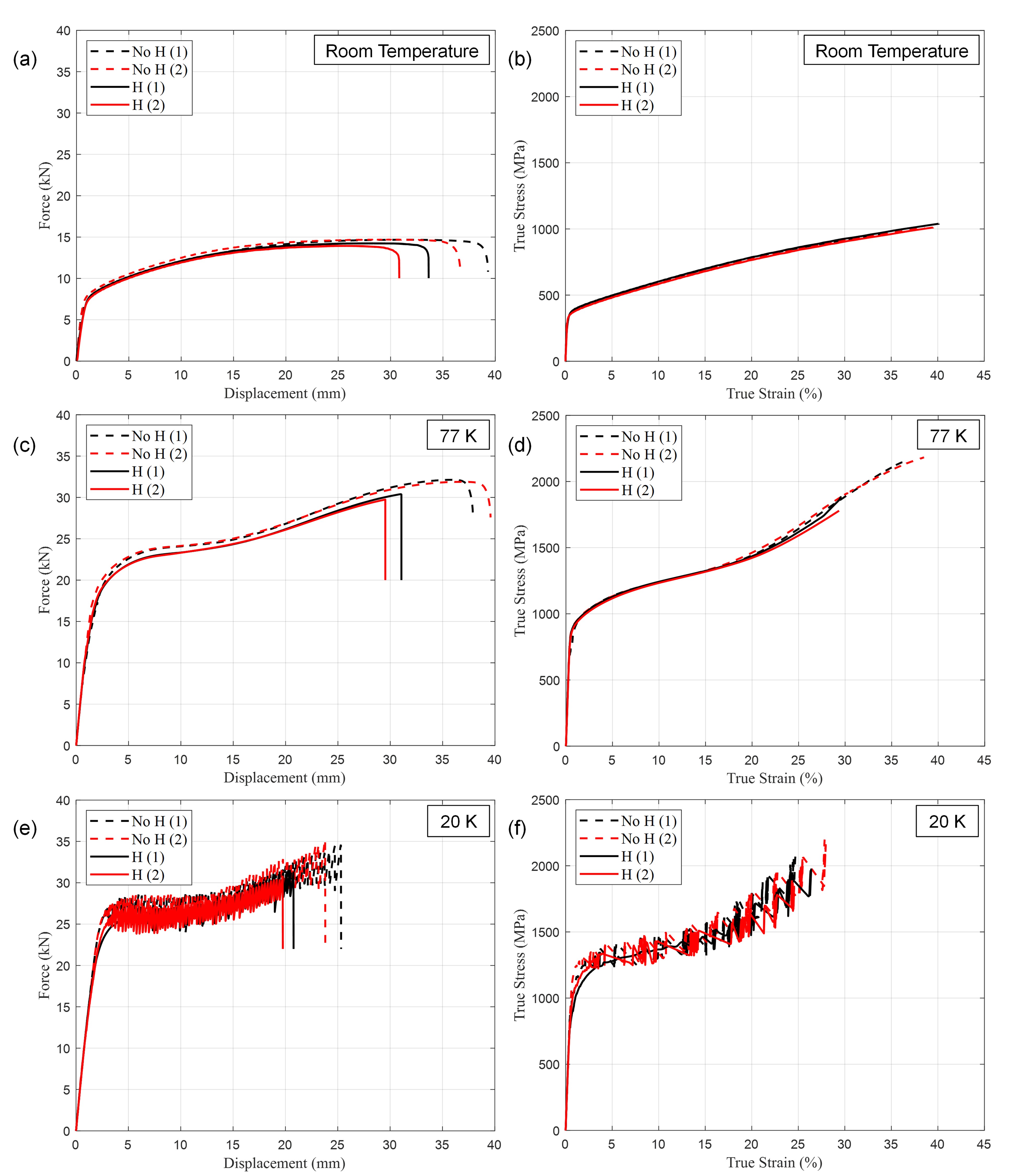}
        \caption{Tensile force–displacement and true stress–strain responses of uncharged and hydrogen pre-charged 316plus specimens tested at room temperature (a,b), 77 K (c,d) and 20 K (e,f). Panels (a,c,e) show the complete force–displacement curves obtained from machine crosshead displacement, while panels (b,d,f) present the corresponding true stress–strain responses measured using a clip-on extensometer. The results illustrate the increase in strength and reduction in ductility with decreasing temperature, as well as the additional degradation associated with hydrogen pre-charging.}
        \label{fig:Figure_5}
\end{figure}

The measured true stress–strain ($\sigma_\mathrm{t}$–$\varepsilon_\mathrm{t}$) responses of uncharged and hydrogen pre-charged specimens at room temperature (RT), 77 K and 20 K are shown in Fig. \ref{fig:Figure_5}. At room temperature, the true stress–strain curves are plotted up to the onset of necking, identified using Considère’s criterion, where the true stress $\sigma_\mathrm{t}$ equals the strain hardening rate $\mathrm{d}\sigma_\mathrm{t} / \mathrm{d}\varepsilon_\mathrm{t}$ \cite{10_bauchau2008structural}. For the cryogenic tests at 77 K and 20 K, the curves are shown up to the point where reliable extensometer data were available. Force-displacement curves, based on machine crosshead displacement, are also included in Fig. \ref{fig:Figure_5} to capture the complete deformation behaviour. Both the stress-strain and force-displacement results exhibited excellent repeatability across all test conditions, with duplicate experiments confirming the reliability of the mechanical response under varying temperature and hydrogen charging states.

%Figure 5 - Stress-strain curves

\begin{figure}[H]
    \centering
    \includegraphics[width=0.6\linewidth]{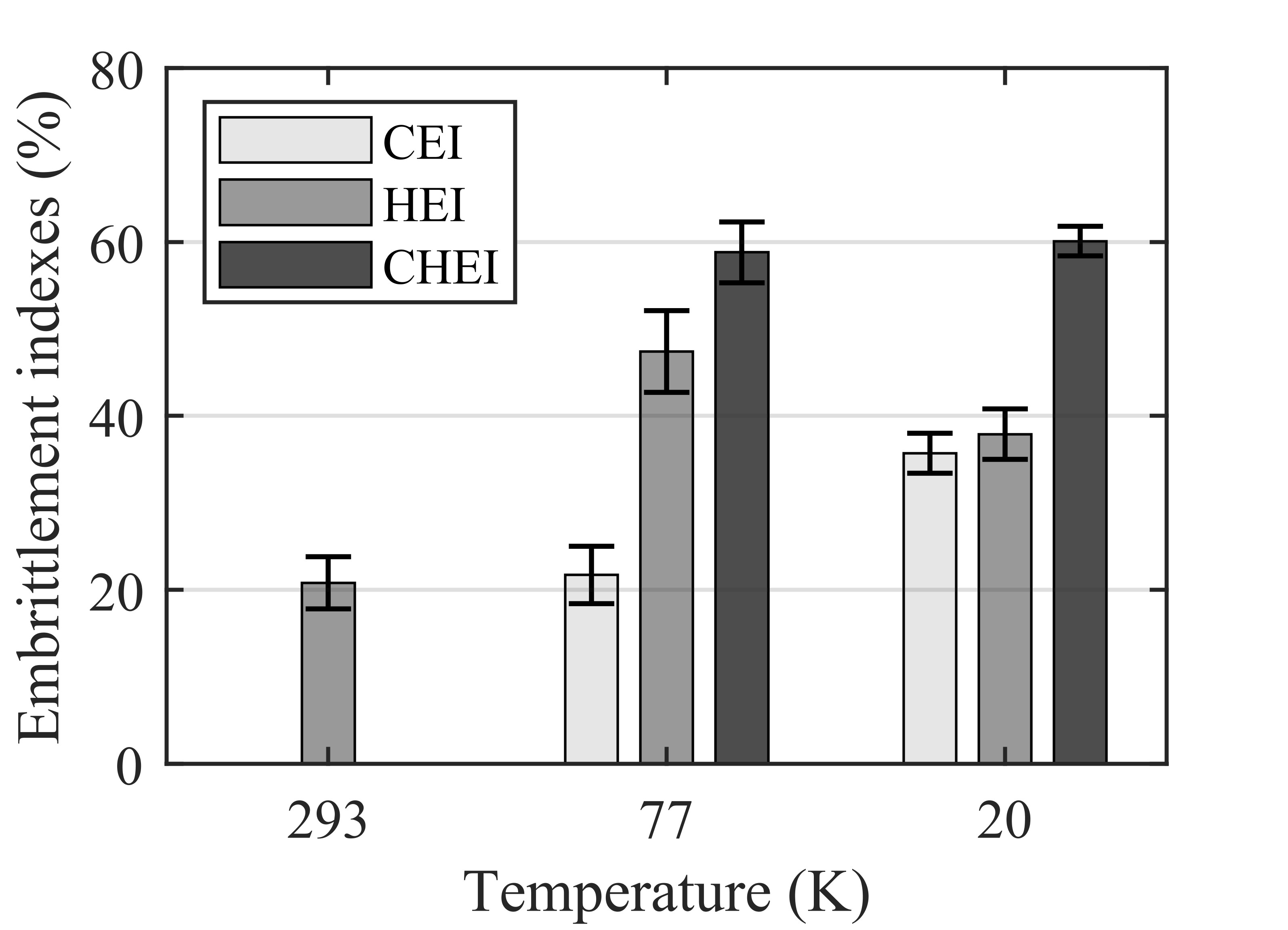}
    \caption{Cryogenic embrittlement index (CEI), hydrogen embrittlement index (HEI), and combined cryogenic–hydrogen embrittlement index (CHEI) for 316plus at room temperature (295 K), 77 K and 20 K. The indexes are calculated from the reduction in area (RA) values reported in Table 2 and quantify the relative contributions of hydrogen charging, cryogenic temperature and their combined effect to ductility loss.}
    \label{fig:CHEI}
\end{figure}

A summary of the key tensile parameters is reported in Table \ref{Table_2}, where $f_\mathrm{y}$ is the yield stress (0.2\% proof stress), $f_\mathrm{u}$ is the ultimate tensile strength (defined as the maximum force divided by the initial cross-sectional area), RA is the reduction in area, and CEI, HEI and CHEI denote the cryogenic, hydrogen and combined hydrogen–cryogenic embrittlement indexes, respectively. The three embrittlement indexes are also plotted in Fig. \ref{fig:CHEI}.

\begin{table}[H]
 \centering
  \caption{Mechanical properties of uncharged and hydrogen pre-charged 316plus specimens at room temperature, 77 K and 20 K.}
  \label{Table_2}
%\resizebox{\textwidth}{!}{
\begin{tabular}[t]{lcccccc}
\toprule
\text{Condition} & $f_{y}$(MPa) & $f_{u}$(MPa) & RA(\%) & CEI or HEI (\%) & CHEI(\%)\\
\midrule
\text{RT-No H} & 341$\pm$2 & 685$\pm$1 & 73.5$\pm$1.6 & - & - \\
\text{77K-No H} & 815$\pm$89 & 1492$\pm$8 & 57.6$\pm$2.1 & 21.7 & -\\
\text{20K-No H} & 1060$\pm$120 & 1644$\pm$29 & 47.3$\pm$1.3 & 35.7 & -\\
\midrule
\text{RT-H} & 341$\pm$4 & 689$\pm$11 & 58.2$\pm$1.8 & 20.8 & -\\
\text{77K-H} & 864$\pm$11 & 1470$\pm$22 & 30.3$\pm$2.5 & 47.4 & 58.8\\
\text{20K-H} & 906$\pm$69 & 1499$\pm$23 & 29.4$\pm$1.1 & 37.9 & 60.1\\
\bottomrule
\end{tabular}
\end{table}

Examination of the stress-strain curves reveals a significant influence of temperature on the mechanical response. At room temperature, the curves display a smooth, nonlinear behaviour without a distinct yield point, followed by gradual strain hardening. The $f_\mathrm{y}$ and $f_\mathrm{u}$ values are highly consistent with those reported by the material supplier, differing less than 1.3\%, which confirms excellent consistency and material uniformity. This overall behaviour is retained in the hydrogen pre-charged specimens; however, a reduction in ductility is evident, as indicated by a $\sim$21\% decrease in the RA (Fig. \ref{fig:CHEI}). In contrast, at 77 K and 20 K, the stress-strain responses of both uncharged and hydrogen pre-charged specimens exhibit an initial linear region up to the yield point, followed by a yield plateau and subsequent strain hardening until fracture. Consequently, the uncharged specimens tested at both cryogenic temperatures showed higher yield and ultimate tensile strengths but lower ductility compared with those tested at room temperature, as indicated by increased CEI (see Fig. \ref{fig:CHEI}). As discussed later, this behaviour arises from the combined effects of low temperature and plastic strain-induced martensitic transformation \cite{11_zheng2018effect}. 

Hydrogen charging further degraded the mechanical performance at low temperatures. Although the overall shape of the stress-strain curves remained comparable to that of the uncharged specimens, the hydrogen pre-charged samples exhibited reduced uniform strain and lower ultimate tensile strength. The RA decreased by approximately $\sim$47\% at 77 K and 38\% at 20 K, indicating an increased susceptibility to hydrogen embrittlement at lower temperatures (see Fig. \ref{fig:CHEI}). At 20 K, serrated flow was observed in both the stress-strain and force-displacement curves for both uncharged and hydrogen pre-charged conditions, particularly in the post-yield region. This serrated behaviour is indicative of discontinuous plastic deformation and has been reported in previous studies of metastable austenitic stainless steels tested at cryogenic temperatures of 20 K or below \cite{Zheng2022low,16_kim2023tensile,17_fernandezpison2021flow,WADA2025}. As discussed later, this phenomenon can be associated with the formation of localised shear bands and dislocation localisation, leading to intermittent stress drops and recoveries.

\subsection{Fracture surface analysis}
\label{Subsec: Fracture surfaces}
\begin{figure}[H]
     \centering
        \centering
    \includegraphics[width=1\textwidth]{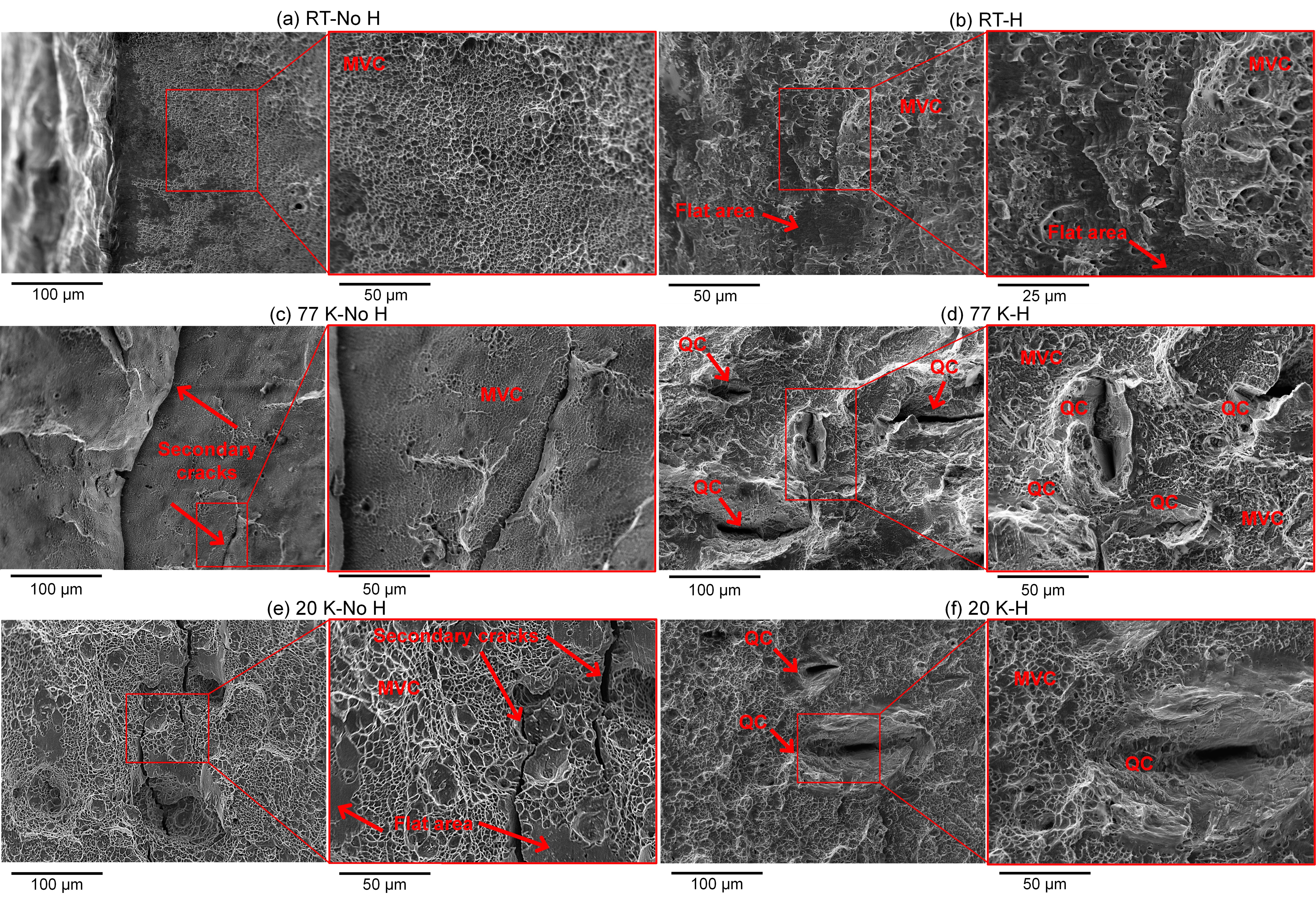}
        \caption{SEM fractographs of tensile fracture surfaces for uncharged and hydrogen pre-charged specimens tested at room temperature (a,b), 77K (c,d) and 20K (e,f). Images correspond to regions located approximately 50–100 µm below the specimen surface, where hydrogen concentrations were estimated to range between $\sim$30–40 wppm. Representative fracture features are indicated with red arrows and labels. MVC=micro-void coalescence; QC=quasi-cleavage.}
        \label{fig:Figure_6}
\end{figure}

The fracture surfaces of specimens tested under all conditions were systematically examined using SEM. Fig. \ref{fig:Figure_6} shows representative micrographs taken at the specimen mid-plane, approximately 50–100 $\upmu$m below the surface - where hydrogen charging was most pronounced and concentrations were estimated to range between 40 and 30 wppm (see Fig. \ref{fig:Figure_3}(b)). These images illustrate the evolution of fracture morphology across the different conditions, ranging from ductile to brittle and mixed-mode failure. At room temperature (Fig. \ref{fig:Figure_6}(a)-(b)), both uncharged and hydrogen pre-charged specimens exhibit predominantly ductile fracture features characterised by extensive micro-void coalescence (MVC). The uncharged specimen displays a uniform network of fine, closely spaced dimples, indicative of significant plastic deformation. In contrast, the hydrogen pre-charged specimen presents a more irregular surface morphology, with larger and shallower dimples and flat regions lacking evidence of micro-void coalescence - suggesting localised suppression of ductile fracture mechanisms. 

%Figure 6 - SEM results

At 77 K (Fig. \ref{fig:Figure_6}(c)-(d)), the fracture surfaces of the uncharged specimens remain mainly ductile, showing small dimples, a relatively smooth topography and evidence of localised plastic deformation. Some secondary cracks are visible (highlighted with red arrows), although the overall behaviour remains consistent with ductile failure. In contrast, the hydrogen pre-charged specimens exhibit a markedly different morphology, with pronounced brittle features. The fracture surface appears more irregular and jagged, with extensive secondary cracking (indicated with red arrows) containing quasi-cleavage (QC) facets, indicating a transition to a more brittle fracture mode under the combined effects of hydrogen and cryogenic temperature, what is referred here as \emph{cryogenic hydrogen embrittlement}. 

At 20 K (Fig. \ref{fig:Figure_6}(e)-(f)), the uncharged specimen still displays a predominantly ductile fracture mode, characterised by a fibrous texture and abundant micro-voids interspersed with some flat regions and secondary cracks (as indicated with red arrows). Conversely, the hydrogen-charged specimen reveals a mixed-mode fracture morphology, combining ductile regions with quasi-cleavage facets and secondary cracking, similar to the 77 K hydrogen pre-charged specimens. Notably, in the hydrogen pre-charged condition, although brittle features are present at 20 K, their extent is reduced compared with those observed at 77 K, where MVC was less apparent - showing shallower and smaller dimples - at the employed magnifications, in line with the measured hydrogen embrittlement index (HEI, Table \ref{Table_2}). 

This trend is consistent with previous studies on austenitic stainless steels of the 304, 310 and 316 series, which exhibit maximum hydrogen embrittlement between 150-200 K, followed by a gradual decrease at lower temperatures \cite{Yang2023, Fukuyama2003}. However, the temperature of maximum embrittlement and the rate of this transition vary significantly with alloy type and grade, reflecting differences in composition, microstructure and hydrogen transport characteristics. As this behaviour arises from the interplay between strain-induced martensitic transformation and hydrogen diffusivity, there is a need to shed light on the mechanisms underlying the behaviour of 316plus in hydrogen and cryogenic temperatures, as required to evaluate the alloy’s suitability for cryogenic applications.

\subsection{Strain-induced martensite}

\begin{figure}[H]
     \centering
    \includegraphics[width=0.85\textwidth]{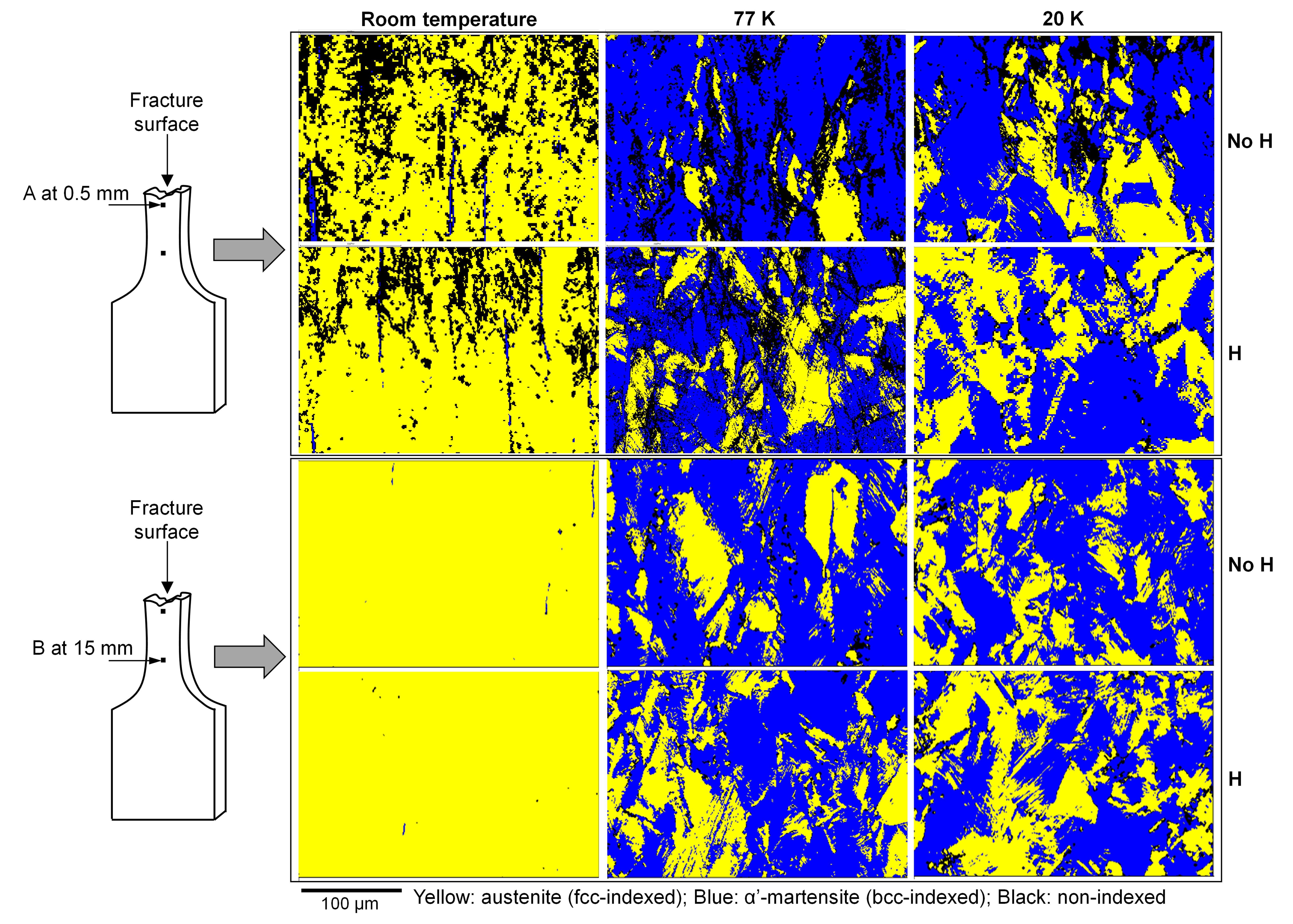}
        \caption{EBSD phase maps of uncharged and hydrogen pre-charged 316plus specimens tested at room temperature, 77 K and 20 K, obtained at two locations along the gauge length: A (0.5 mm) and B (15 mm) from the fracture surface. Austenite (fcc) is shown in yellow and strain-induced martensite (bcc) is shown in blue. The maps illustrate the progressive increase in strain-induced martensite formation at cryogenic temperatures and the potential influence of hydrogen on the extent of the $\gamma \rightarrow \alpha'$ transformation.}
        \label{fig:Figure_7}
\end{figure}

To characterise the temperature, strain-, and hydrogen-dependent phase transformation behaviour, EBSD phase analysis was carried out on longitudinal cross-sections of fractured specimens at approximately 0.5 mm and 15 mm from the fracture surface, and at a depth of $\sim$100 $\upmu$m below the surface. This depth corresponds to the region examined in the SEM fractography, where hydrogen concentrations were estimated to be 30-40 wppm. Fig. \ref{fig:Figure_7} presents EBSD phase maps for uncharged and hydrogen pre-charged specimens tested at RT, 77 K and 20 K at these two locations - A (0.5 mm) and B (15 mm) from the fracture. Austenite (fcc) is shown in yellow, and strain-induced martensite (bcc) in blue; black pixels correspond to non-indexed regions, seen particularly near the fracture at room temperature, where severe deformation produced substantial lattice distortion. 

%Figure 7 - EBSD results

Because the formation of $\alpha'$-martensite (and thus EBSD phase contrast) is highly sensitive to local plastic strain, the phase fraction data were evaluated together with the true plastic strain ($\varepsilon_\mathrm{t}$) measured at locations A and B using initial and current cross-sectional areas. Local strain was further assessed using Kernel Average Misorientation (KAM) maps (see \ref{Appendix:A}), enabling a more rigorous comparison of SIM fractions across different tested conditions. The true strain, mean KAM values and measured bcc and fcc phase fractions are summarised in Table \ref{tab:SIM_KAM}. Relative SIM fractions (SIM$\mathrm{_{rel}}$) were calculated using only indexed points, excluding non-indexed pixels, which provides the most reliable representation of phase fractions in EBSD datasets \cite{Niessen2021}. The KAM values follow the same trend as the true strain measurements, confirming that both metrics capture the local deformation state along the gauge length.

\begin{table}[H]
\centering
\caption{Summary of true strain, mean KAM and relative strain-induced martensite (SIM$\mathrm{_{rel}}$) fractions at two locations for all test conditions.}
\begin{tabular}{clccccc}
\hline
\text{Location} & \text{Condition} & \text{True strain} & \text{Mean KAM} & \text{bcc (\%)} & \text{fcc (\%)} & \text{SIM$\mathrm{_{rel}}$ (\%)} \\
\hline
& RT- No H & 0.72 & 2.02 & 1.0 & 78.9 & 1.3 \\
& RT- H  & 0.70 & 1.72 & 0.6 & 83.7 & 0.7 \\
A & 77K- No H & 0.63 & 1.92 & 67.3 & 9.5 & 87.6  \\
(0.5 mm) & 77K- H & 0.25 & 1.35 & 53.6 & 28.5 & 65.3 \\
& 20K- No H & 0.22 & 1.37 & 56.9 & 26.6 & 68.1 \\
& 20K- H & 0.20 & 1.47 & 56.1 & 40.5 & 58.1 \\
\midrule
& RT- No H & 0.40 & 1.87 & 0.16 & 99.8 & 0.2 \\
& RT- H & 0.36 & 1.44 & 0.03 & 99.9 & 0.03 \\
B & 77K- No H & 0.35 & 1.76 & 69.3 & 25.4 & 73.2 \\
(1.5 mm) & 77K- H & 0.25 & 1.39 & 61.1 & 34.6 & 64.0 \\
& 20K- No H & 0.20 & 1.19 & 65.0 & 31.5 & 67.4 \\
& 20K- H & 0.19 & 1.21 & 52.8 & 41.3 & 56.1 \\
\hline
\end{tabular}
\label{tab:SIM_KAM}
\end{table}

At room temperature, the microstructure remained almost fully austenitic for both uncharged and hydrogen pre-charged specimens at both the high-strain (A) and low-strain (B) locations. Only minimal SIM was detected (SIM$\mathrm{_{rel}}$ $<$ 1.3\% at A and $<$ 0.2\% at B), in agreement with the high austenite stability of 316plus (with $M_{\mathrm{d}}^{30}$=\(-124\ ^{\circ}\mathrm{C}\) and \(M_\mathrm{s}\)=\(-285\ ^{\circ}\mathrm{C}\)). The small differences in SIM between uncharged and hydrogen pre-charged specimens were within expected EBSD variability and consistent with reported behaviour for 316L \cite{KOMATSU2021,sanmachi2021}. At cryogenic temperatures (77 K and 20 K), a substantial increase in SIM was observed in both uncharged and hydrogen pre-charged specimens. At the more strained location A, SIM$\mathrm{_{rel}}$ reached 58-88\%, while at location B it reached 56-73\%. The EBSD maps show a clear dominance of the bcc phase at both temperatures, demonstrating the strong effect of low temperature on the $\gamma \rightarrow \alpha'$ transformation. This is consistent with the increase in flow stress and strain-hardening seen in the mechanical response (Fig. \ref{fig:Figure_5}).

SIM increased progressively as the temperature decreased from RT to cryogenic levels, reaching 58-88\% at 77 K and 20 K (Table \ref{tab:SIM_KAM}). At room temperatures, hydrogen appears to have only a minor influence on the SIM fraction. At 77 K, however, interpretation of the hydrogen effect is complicated by the different local strain levels in hydrogen pre-charged and uncharged specimens (Table \ref{tab:SIM_KAM}). Since SIM formation is strongly strain-dependent, the lower SIM fraction in the hydrogen pre-charged condition cannot be attributed to hydrogen alone. Therefore, the 77 K dataset does not provide a strain-matched basis for isolating the hydrogen contribution to the transformation. At 20 K, however, both strain level metrics (true strain and KAM) were comparable between hydrogen pre-charged and uncharged specimens at both locations A and B. Under these conditions, the uncharged specimen exhibited higher SIM fractions (e.g. SIM$\mathrm{_{rel}}$ = 67.4\% vs. 56.1\%), suggesting that hydrogen may reduce the extent of the $\gamma \rightarrow \alpha'$ transformation at this temperature. A detailed interpretation of these results, including mechanistic implications and comparison with 316L at different strain levels and hydrogen levels, is provided in the Discussion section.

%%%%%%%%%%%%%%%%%%%%%%%%%% DISCUSION %%%%%%%%%%%%%%%%%%%%%%%%%%%%%%%

\section{Discussion}
\subsection{Stress-strain response under cryogenic and hydrogen conditions}

\begin{figure}[H]
     \centering
         \centering
         \includegraphics[width=1\textwidth]{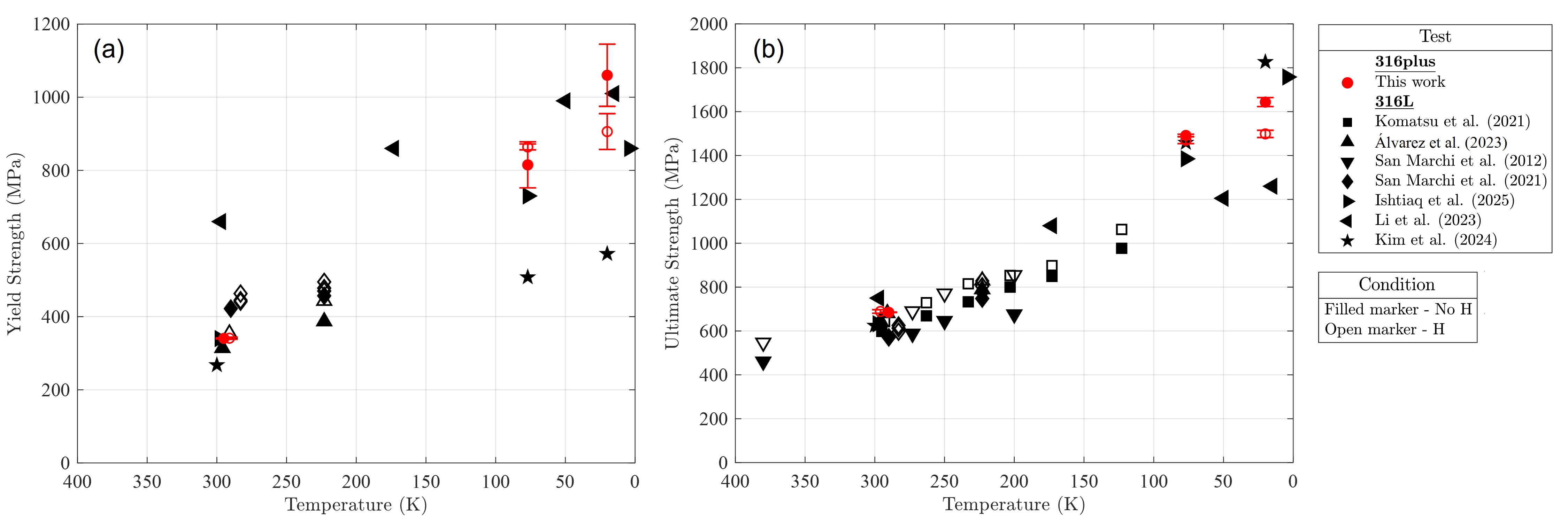}
        \caption{Comparison of (a) yield strength and (b) ultimate tensile strength of 316plus measured in this work with literature data for 316L \cite{KOMATSU2021,8_alvarez2023hydrogen,sanmachi2021,SanMarchi2012,ISHTIAQ2025,Li2023,KIM2024} under uncharged (filled markers) and hydrogen pre-charged (open markers) conditions across the temperature range. The results illustrate the progressive increase in strength with decreasing temperature and show that 316plus exhibits strength values at the upper bound of the reported 316L data, while hydrogen has only a minor effect on the strength parameters.}
        \label{fig:Figure_9}
\end{figure}

Figs \ref{fig:Figure_9}(a) and (b) show the evolution of yield strength ($f_\mathrm{y}$) and ultimate strength ($f_\mathrm{u}$) with temperature for the tested 316plus stainless steel, together with literature data for 316L \cite{KOMATSU2021,8_alvarez2023hydrogen,sanmachi2021,SanMarchi2012,ISHTIAQ2025,Li2023,KIM2024} under both uncharged and hydrogen pre-charged conditions. The literature data originate from materials produced using a range of processing routes; however, most correspond to hot-rolled products and exhibit broadly consistent trends with temperature and hydrogen. The pre-charged dataset enables a direct comparison with the present results, and is particularly relevant because pre-charging provides a clearer measure of the effect of hydrogen than in-situ testing, owing to the much lower hydrogen diffusivity during cryogenic tensile loading. As 316plus is a nitrogen-alloyed derivative of 316L, sharing a similar austenitic matrix but offering higher strength and corrosion resistance, direct comparison with 316L provides a meaningful reference for assessing its mechanical response. Both yield and ultimate tensile strengths increase markedly with decreasing temperature, with 316plus consistently positioned at the upper bound of the literature trend for 316L at 77 K and 20 K, demonstrating slightly superior strength at low temperature. In agreement with previous works, hydrogen had little ($<$ 8\%) to no effect on either yield or ultimate strength at temperatures above 77 K. However, at 20 K - where no comparable 316L data are available - hydrogen caused a modest reduction of approximately 15\% in the yield strength and 10\% in ultimate strength.  

%Figure 9 - Trends

The mechanical response at 20 K is particularly noteworthy, as this is the temperature at which serrated flow was observed in the tensile stress–strain curves for both uncharged and hydrogen pre-charged specimens (Fig. \ref{fig:Figure_5}). The serrations appeared in the post-yield plastic region and are consistent with previous observations for 316L stainless steel tested between 0–35 K in the absence of hydrogen \cite{ISHTIAQ2025,Obst1991}. At 77 K, by contrast, deformation proceeded smoothly without serrated flow, consistent with earlier findings that place the onset of cryogenic discontinuous plastic flow in austenitic stainless steels at temperatures below approximately 35 K \cite{Tabin2019}. Serrated flow in this low temperature range (0-35 K) has been associated to both phase transformation phenomena and dislocation-microstructure interactions. In an early work, Obst and Nyilas \cite{Obst1991} attributed the serrated behaviour to a mechanical origin, whereby edge-dislocation pile-ups generate local stress concentrations that trigger intermittent bursts of dislocation motion, producing stepwise strain increases and abrupt stress drops. More recently, Ishtiaqi et al. \cite{ISHTIAQ2025} reported serrated flow in 316L stainless steel at 4.2 K, where dislocation pile-ups at twin boundaries within strain-induced martensite acted as barriers to slip. Once the applied stress reached a critical value, these barriers collapsed, leading to avalanche-like dislocation motion and observed serrations. 

In the present study, similar mechanisms are likely responsible for the serrated behaviour observed at 20 K in both uncharged and hydrogen pre-charged specimens. The comparable serrated patterns under both conditions suggest that hydrogen does not significantly alter the mechanism governing serrated deformation at this temperature. This observation is consistent with the findings of Harris et al. \cite{HARRIS2018}, who demonstrated in polycrystalline nickel that mobile hydrogen–deformation interactions are effectively suppressed at cryogenic temperatures, with hydrogen already segregated at grain boundaries prior to deformation playing a dominant role. By analogy, the negligible difference in serrated behaviour between uncharged and hydrogen pre-charged 316plus specimens at 20 K indicates that the contribution of mobile hydrogen to the serrated response is limited under such low-diffusivity conditions. Instead, any hydrogen-related effects are more plausibly associated with hydrogen trapped at microstructural features prior to testing, while the serrated flow itself primarily reflects intrinsic low-temperature deformation mechanisms.

\subsection{Ductility loss and fracture mechanisms under cryogenic and hydrogen conditions}

\begin{figure}[H]
     \centering
         \centering
         \includegraphics[width=0.75\textwidth]{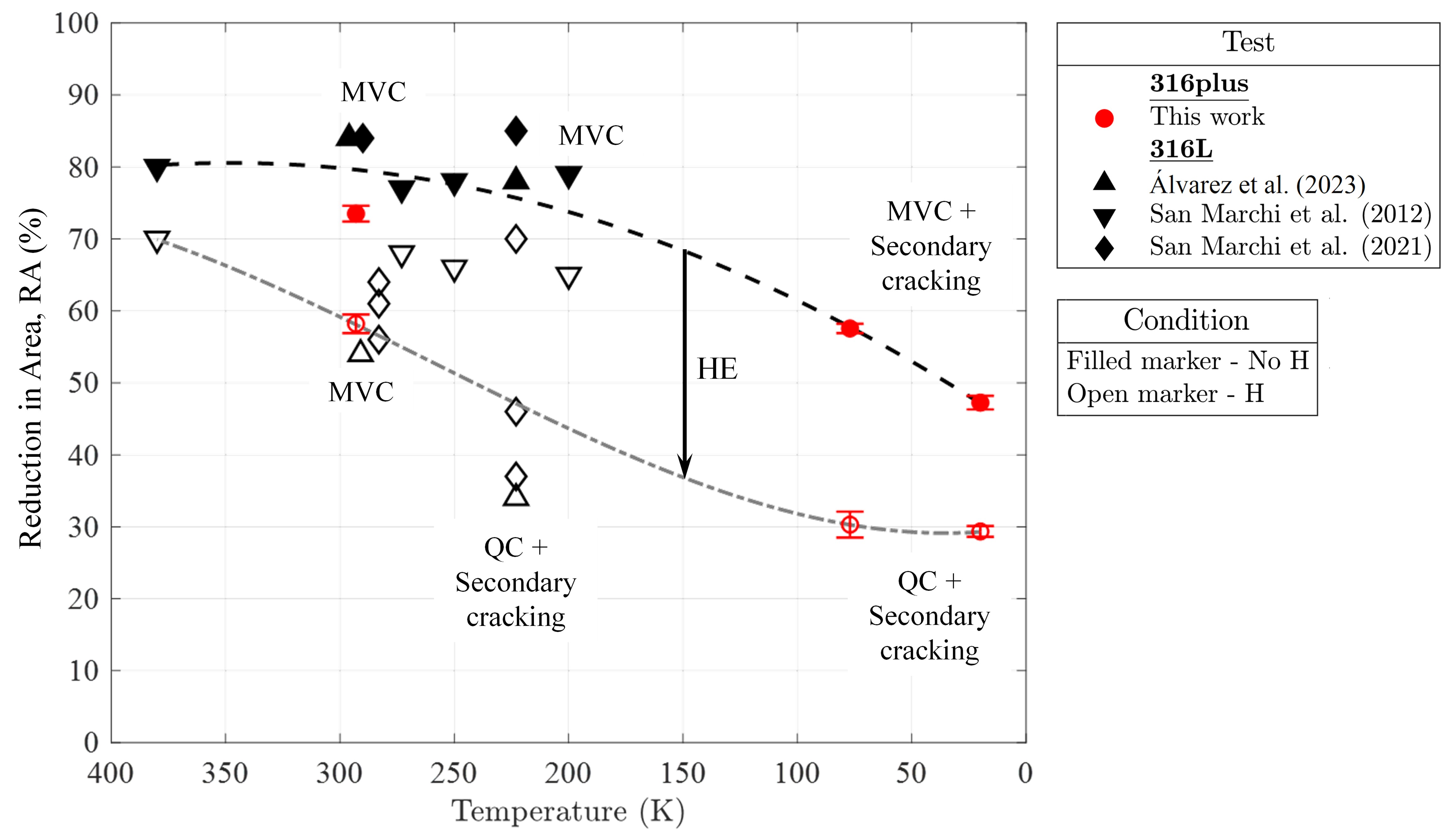}
        \caption{Evolution of reduction in area (RA) with temperature for 316plus (this work) and 316L reported in the literature \cite{8_alvarez2023hydrogen, SanMarchi2012, sanmachi2021} under uncharged (filled markers) and hydrogen pre-charged (open markers) conditions. Trend lines are included for illustration purposes. The literature data correspond to specimens with uniform hydrogen concentrations in the range of 50–140 wppm.}
        \label{fig:Figure_10}
\end{figure}

Fig. \ref{fig:Figure_10} shows the measured reduction in area (RA) values for 316plus tested at RT, 77 K and 20 K, together with literature data for 316L \cite{8_alvarez2023hydrogen, SanMarchi2012, sanmachi2021} over the temperature range 375-200 K, in both uncharged and hydrogen pre-charged conditions. In the absence of hydrogen, the combined dataset for 316plus and 316L indicates that this family of alloys retain high ductility over a broad temperature range, with RA remaining nearly constant at approximately 80\% down to 200 K, and with microvoid coalescence (MVC) being the dominant fracture micromechanism in this regime, consistent with previous studies \cite{8_alvarez2023hydrogen}. At lower temperatures - where only the present 316plus data are available - RA decreases gradually with decreasing temperature, reaching values slightly below 50\% at 20 K. The fracture surfaces of specimens tested at 77 K and 20 K, though more brittle in appearance, still show evidence of ductile tearing, but with the additional presence of secondary cracking.

%Figure 10 - Trends

In contrast, the behaviour of hydrogen pre-charged 316L and 316plus differs markedly from that of uncharged specimens. Despite the scatter in published data - largely attributable to variations in charging conditions and hydrogen concentrations - hydrogen pre-charged samples consistently exhibit lower RA values across the entire range. It is interesting to note that this compilation of data reveals that, although the specimens tested in the present work were not charged to full saturation (see concentration profile in Fig. \ref{fig:Figure_3}), the degree of embrittlement measured at room temperature aligns closely with the average value reported in the literature for specimens saturated to 50–140 wppm. This suggests that the selected charging conditions were sufficient to capture the material’s hydrogen-assisted response. Although the hydrogen distribution in the present specimens is non-uniform, the diffusion simulations indicate that hydrogen concentrations approaching $\sim$52 wppm were reached within a $\sim$0.5 mm layer from each surface, representing a substantial fraction of the specimen cross-section. Since fracture initiation and damage accumulation during tensile testing typically occur in the near-surface regions where hydrogen concentration is highest, the measured RA reductions are expected to be governed primarily by these hydrogen-enriched regions. The influence of hydrogen becomes more pronounced with decreasing temperature, appearing to reach its maximum impact between 200 and 100 K. Hydrogen-induced degradation persist down to cryogenic temperatures, although the rate of RA reduction appears to attenuate between 70-20 K. Notably, in the present study, although the absolute RA values of hydrogen pre-charged specimens at 77 K and 20 K are broadly similar, the relative RA loss with respect to uncharged specimens - quantified by the hydrogen embrittlement index (HEI) - is greater at 77 K ($\sim$47\%) than at 20 K ($\sim$40 \%). This indicates a comparatively higher susceptibility to hydrogen embrittlement at 77 K. Overall, the RA trends in Fig. \ref{fig:Figure_10} suggest that the behaviour of uncharged and hydrogen pre-charged specimens begins to converge as the temperature approaches cryogenic conditions. 

Although MVC remained the predominant fracture micromechanism for hydrogen pre-charged specimens tested at room temperature, consistent with comparable works \cite{8_alvarez2023hydrogen}, these samples still exhibited substantial hydrogen embrittlement, with ductility losses exceeding 20\%. At cryogenic temperatures (77 K and 20 K), the fracture mode shifted sharply towards brittle behaviour. While traces of MVC were still present, the fracture surfaces were dominated by quasi-cleavage facets and extensive secondary cracking, accompanied by RA values reduced to 40-50\%. For 316L tested at intermediate temperatures (e.g., 223 K) other authors have similarly reported cleavage-like features and significant secondary cracking \cite{8_alvarez2023hydrogen}, demonstrating that the emergence of brittle mechanisms does not require reaching temperatures as low as 70 K. 

This temperature-dependent change in fracture behaviour may be linked to the progressive reduction in hydrogen diffusivity. As the temperature approaches 20 K, diffusivity becomes extremely limited, restricting the transport of hydrogen to deformation-induced defects and thereby moderating its contribution to embrittlement. A recent meso-scale modelling study by de Melo Freire et al. \cite{deMeloFreire2025} examined the temperature sensitivity of hydrogen diffusion and its implications for hydrogen embrittlement in 316L stainless steel. The authors found that, at moderately low temperatures, strain-induced martensitic transformation can locally increase hydrogen diffusivity, enabling more rapid hydrogen transport during plastic deformation and thereby enhancing susceptibility to embrittlement. However, at lower temperatures (below $\sim$198 K), hydrogen diffusion becomes increasingly restricted. It should be noted that at very low temperatures, hydrogen transport may deviate from classical Arrhenius behaviour due to quantum tunnelling effects, which can lead to diffusivities higher than those predicted by high-temperature extrapolation \cite{KIRCHHEIM2004}. Nevertheless, even when tunnelling is considered, hydrogen mobility in austenitic stainless steels remains extremely limited under cryogenic conditions. Their model therefore suggests the existence of a temperature regime in which hydrogen transport becomes strongly constrained, leading to a reduced hydrogen effect at cryogenic temperatures. Experimentally validating this behaviour remains challenging due to the scarcity of low-temperature diffusivity data and the inherent difficulties in quantifying hydrogen mobility under such conditions. Nevertheless, the present results appear consistent with this interpretation. The modelling work also highlighted the central role of strain-induced martensitic transformation in governing hydrogen embrittlement in metastable austenitic stainless steels within specific temperature ranges - an aspect that will be examined in detail in the following section.

\subsection{Influence of temperature and hydrogen on strain-induced martensite formation}

\begin{figure}[H]
    \centering
    \includegraphics[width=0.7\linewidth]{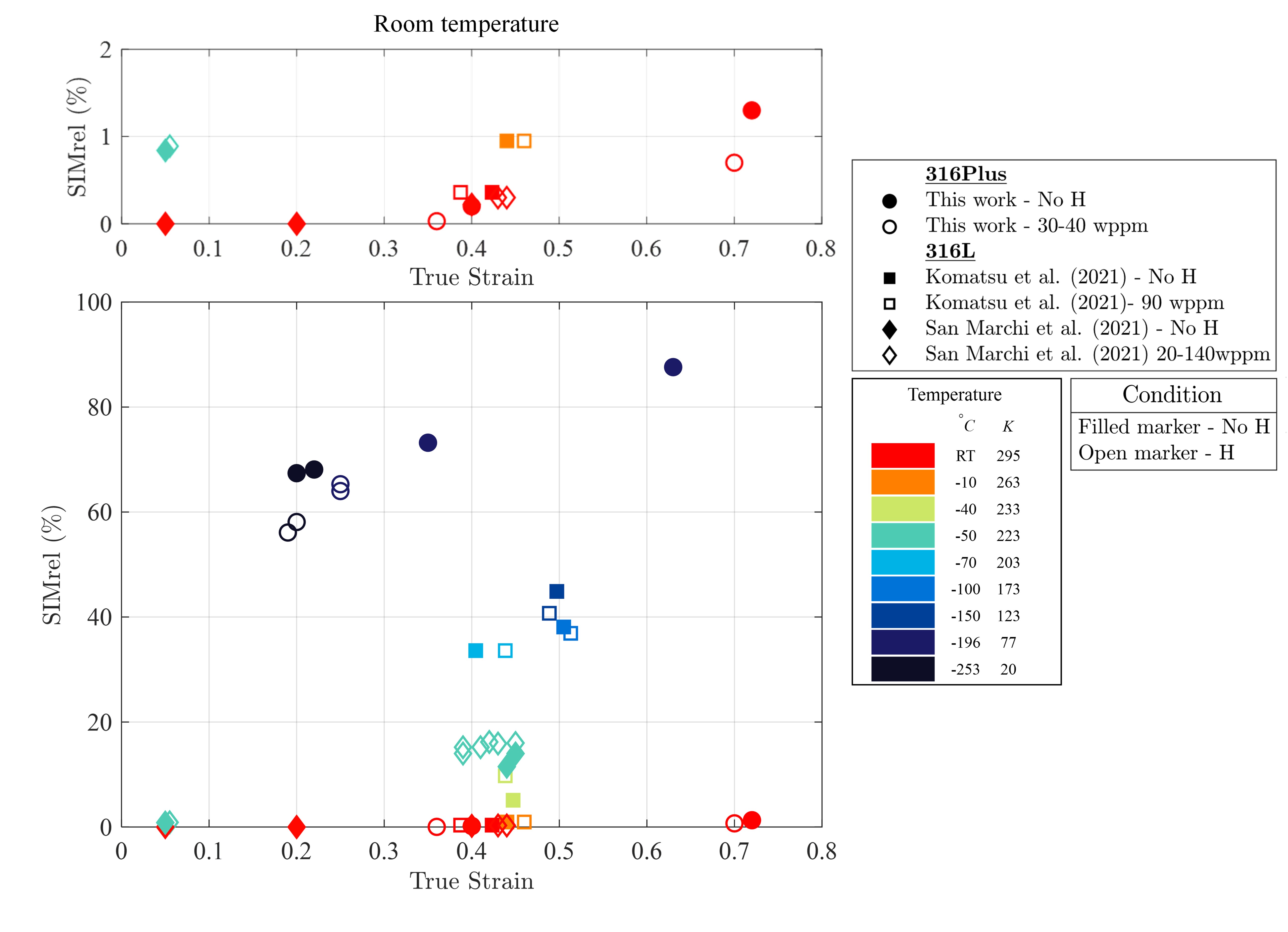}
    \caption{Evolution of strain-induced martensite (SIM$_\mathrm{rel}$) as a function of true plastic strain and temperature for 316plus, together with literature data for 316L \cite{sanmachi2021, KOMATSU2021}. Filled markers correspond to uncharged specimens and open markers to hydrogen pre-charged specimens. The data illustrate the increase in SIM formation with plastic strain and decreasing temperature, as well as the comparatively limited influence of hydrogen on the overall martensitic transformation behaviour.}
    \label{fig:Figure_11}
\end{figure}

The formation of strained-induced $\alpha'$-martensite in austenitic stainless steels is governed by alloy composition, accumulated plastic strain, deformation temperature and the presence of hydrogen \cite{sanmachi2008}. To interpret the present results in this mechanistic context - and to enable comparison with the most closely related alloy, 316L - Fig. \ref{fig:Figure_11} summarises the evolution of SIM$_\mathrm{rel}$ as a function of true plastic strain and temperature for 316plus, alongside representative datasets for uncharged and hydrogen pre-charged 316L \cite{sanmachi2021, KOMATSU2021}. Whereas literature studies commonly quantify SIM using interrupted tests at controlled strain levels, the 316plus specimens were strained to fracture; SIM was therefore assessed at two locations of the the gauge length with different local strains (Fig. \ref{fig:Figure_7}) and interpreted together with both true strain and KAM-based deformation metrics. 

%Figure 11: SIM and strain

The data in Fig. \ref{fig:Figure_11} highlight three key trends. First, SIM increases monotonically with plastic strain, consistent with the well-established role of dislocation activity, shear bands and defect accumulation in providing nucleation paths for $\alpha'$-martensite in metastable austenite \cite{KOMATSU2021, sanmachi2021}. The present 316plus results follow this behaviour closely: regions near the fracture, where local strain and KAM values are highest, exhibit the largest SIM fractions. Second, decreasing temperature enhances SIM formation. This reflects the reduced stacking-fault energy (SFE) at low temperature, which promotes planar slip and mechanically assisted $\gamma \rightarrow \alpha'$ transformation \cite{11_zheng2018effect,13_ding2019modified,14_ding2018tensile}. Composition-based estimates of SFE measures for 316plus (34 mJ/m$^2$ at RT, 23 mJ/m$^2$ at 77 K and 20 mJ/m$^2$ at 20 K \cite{ISHTIAQ2025}) are consistent with the greatly increased SIM fractions observed at both cryogenic temperatures. Although no SIM data for 316L exist below 75 K, the low-temperature transformation behaviour observed here aligns well with the expected thermodynamic trend for 316-series steels. Third, and most critically, the influence of hydrogen on SIM formation depends strongly on temperature.

At room and relatively high temperatures (295-263 K), hydrogen appears to have a negligible effect on SIM formation, consistent with the small differences observed between uncharged and hydrogen pre-charged 316 plus at RT and with reported behaviour for 316L. At intermediate low temperatures ($\approx$233-223 K in the literature), hydrogen tends to slightly promote the nucleation of SIM at low transformation fractions ($<$20\%), a behaviour attributed to hydrogen-enhanced planar slip and intensified shear-band intersections \cite{sanmachi2021}. By contrast, at higher transformation fractions and lower temperatures (173 K and below) - the regime probed in the present study - hydrogen suppresses further $\alpha'$-martensite growth. This trend is reproduced clearly in the 20 K 316plus data: under strain-matched conditions, hydrogen pre-charged specimens contain 10-15\% less $\alpha'$-martensite than uncharged specimens at both examined locations. This suppression is consistent with mechanistic interpretations proposed for 316L, where hydrogen is argued to stabilise the austenite by promoting planar slip, reducing cross-slip and hindering the development of the dislocation configurations required for SIM thickening \cite{sanmachi2021, Bak2017}, even though those studies were carried out at higher temperatures. The observation that hydrogen concentration (20–120 wppm in the literature) produces only modest differences in SIM fractions further supports the interpretation that hydrogen alters deformation mode more than transformation kinetics.

A key implication of the combined dataset is that the absolute SIM fraction does not govern hydrogen embrittlement in 316plus. Multiple authors have shown that hydrogen-assisted fracture in austenitic stainless steels is controlled primarily by hydrogen-dislocation interactions, decohesion of critical interfaces, slip localisation and the accumulation of vacancy-type defects at slip-band intersections, rather than by the extent of $\alpha'$-martensite formation itself \cite{ROZENAK1990,sanmachi2008, sanmachi2021}. The present findings reinforce this view: although hydrogen appears to suppress SIM at 20 K, hydrogen-charged specimens nevertheless show markedly reduced ductility and increased brittle fractures at both 77 K and 20 K. This indicates that trapped hydrogen, introduced during pre-charging and retained at cryogenic temperatures, accelerates damage initiation even when the transformed martensite fraction is lower. Conversely, uncharged specimens can exhibit high SIM fractions at 77 K and 20 K while still retaining significantly greater ductility.

Taken together, these results position 316plus squarely within the mechanistic framework established for 316L: temperature primarily dictates the extent of SIM through its influence on austenite stability and SFE, whereas hydrogen modifies deformation and cracking behaviour, promoting planar slip and local interface decohesion. Thus, at cryogenic temperatures, the degree of hydrogen embrittlement in 316plus is not controlled by the amount of SIM but by hydrogen-modified plasticity and micro-damage evolution, with SIM serving as an indicator rather than a driver of the operative deformation pathway. 

%%%%%%%%%%%%%%%%%%%%%%%%%% CONCLUSIONS %%%%%%%%%%%%%%%%%%%%%%%%%%%%%%%
\section{Conclusions}

This study investigated, for the first time, the combined effects of cryogenic temperature and internal hydrogen on the mechanical behaviour and embrittlement susceptibility of a new conventionally rolled austenitic stainless steel, 316plus (EN 1.4420), developed for liquid hydrogen containment. Uniaxial tensile tests were conducted at room temperature, 77K and 20K on uncharged and hydrogen pre-charged specimens, and the results were interpreted together with detailed fractographic analysis and quantification of strain-induced martensite. The main conclusions are:

\begin{itemize}

    \item At 77 K and 20 K, stainless steel 316plus exhibited significantly higher yield and ultimate tensile strengths compared with room temperature, due to the strengthening effect of strain-induced martensite (SIM) transformation, facilitated by the reduced stacking fault energy and lower austenite stability at low temperatures.
    
    \item Yield and ultimate tensile strengths of 316plus were largely unaffected by hydrogen at room temperature and 77 K. A modest reduction (around 10\%) was observed only at 20 K. Across all temperatures, 316plus exhibited strengths at the upper bound of reported values for 316L, demonstrating superior cryogenic performance, despite its lower nominal Ni content, owing to its comparable nickel equivalent ($\mathrm{Ni_{eq}}$) and maintained austenite stability.
    
    \item Hydrogen embrittlement was substantial at all temperatures and reached its maximum at cryogenic temperatures (70-20K), where reductions in area (HEI $\approx$ 40-50 \%) and the elevated cryogenic hydrogen embrittlement index (CHEI) indicated a more brittle fracture mode, accompanied by quasi-cleavage and extensive secondary cracking.
    
    \item At 20 K, hydrogen was found to potentially suppress the formation of strain-induced martensite (SIM) in specimens containing 30-40 wppm of hydrogen. When considered in the context of 316L data from the literature, hydrogen appears to promote SIM at intermediate low temperatures (where SIM fractions are still small) but suppresses SIM at cryogenic temperatures - an effect that is relatively insensitive to hydrogen content within the 20–120 wppm range.

    \item The SIM fraction alone does not control hydrogen embrittlement in 316plus; instead, embrittlement is governed by damage mechanisms associated with trapped hydrogen within narrow deformation zones, which can occur even when SIM formation is suppressed, particularly under cryogenic conditions where hydrogen mobility is negligible.

    \end{itemize}

Overall, although hydrogen embrittlement is significant at cryogenic temperatures (70-20 K), 316plus retained considerable ductility (RA=30\%), with plastic deformation persisting even under the most severe conditions. These findings show that 316plus - despite its reduced Ni content relative to traditional 316L - remains mechanically robust under cryogenic hydrogen exposure and is therefore a promising candidate for liquid-hydrogen storage and containment. This study provides the first experimental characterisation of the hydrogen–temperature interaction in 316plus at cryogenic conditions. 
 
\section*{Acknowledgements}

\noindent The authors acknowledge the financial support from the Royal Academy of Engineering and the Leverhulme Trust Research Fellowship as well as the Institution of Civil Engineers Research and Development Enabling Fund awarded to S. Afshan, and the EPSRC (grants EP/Y037219/1 and EP/V009680/2). E. Mart\'{\i}nez-Pa\~neda was additionally supported by an UKRI Future Leaders Fellowship [grant MR/V024124/1]. Technician support from Institute of Cryogenics, University of Southampton, is acknowledged. The technical support and provision of test materials by Outokumpu are also gratefully acknowledged.

%%\appendix
%%\setcounter{table}{0} % Reset table counter
%%\setcounter{figure}{0} % Reset figure counter
%%\renewcommand\thetable{\Alph{section}.\arabic{table}}  % Label tables as A.1, A.2, etc.
%%\renewcommand\thefigure{\Alph{section}.\arabic{figure}} % Label figures as A.1, A.2, etc.

% Content for Appendix A goes here
%%\section{Local strain - KAM maps}
%%\label{Appendix:A}

%%To assess local strain in the locations analysed with EBSD for the determination of strain-induced martensite fraction, this appendix presents the Kernel Average Misorientation (KAM) maps for uncharged and hydrogen-charged specimens at different temperatures (295 K, 70 K and 20 K) at two different locations (A=0.5 mm and B=1.5 mm) from the fracture surface - see Fig. \ref{fig:FigA1_KAMmaps}. 

%%\begin{figure}[H]
%%     \centering
%%         \centering
%%         \includegraphics[width=1\textwidth]{Figure_14.jpg}
%%        \caption{KAM maps in two different locations of the gauge region for uncharged and hydrogen pre-charged samples tested at different temperatures (295 K, 70 K, 20 K).}
%%        \label{fig:FigA1_KAMmaps}
%%\end{figure}

%% If you have bibdatabase file and want bibtex to generate the
%% bibitems, please use
%%
%%  \bibliography{<your bibdatabase>}

\small
\begin{thebibliography}{10}
\expandafter\ifx\csname url\endcsname\relax
  \def\url#1{\texttt{#1}}\fi
\expandafter\ifx\csname urlprefix\endcsname\relax\def\urlprefix{URL }\fi
\expandafter\ifx\csname href\endcsname\relax
  \def\href#1#2{#2} \def\path#1{#1}\fi

\bibitem{suwaileh2025exploring}
W.~Suwaileh, Y.~Bicer, S.~Al~Hail, S.~Farooq, R.~M. Yunus, N.~N. Rosman,
  I.~Karajagi, Exploring hydrogen fuel as a sustainable solution for
  zero-emission aviation: Production, storage, and engine adaptation
  challenges, International Journal of Hydrogen Energy 121 (2025) 304--325.
\newblock \href
  {https://doi.org/https://doi.org/10.1016/j.ijhydene.2025.03.348}
  {\path{doi:https://doi.org/10.1016/j.ijhydene.2025.03.348}}.

\bibitem{1_imo2023ghgstrategy}
{International Maritime Organization}, 2023 {IMO} strategy on reduction of
  {GHG} emissions from ships (2023).

\bibitem{cardenas2025physical}
B.~Cardenas, S.~D. Garvey, L.~Swinfen-Styles, Z.~Baniamerian, C.~N. Eastwick,
  D.~Grant, The physical exergy in hydrogen--maximising the utility of hydrogen
  as an aviation fuel, International Journal of Hydrogen Energy 177 (2025)
  151594.
\newblock \href
  {https://doi.org/https://doi.org/10.1016/j.ijhydene.2025.151594}
  {\path{doi:https://doi.org/10.1016/j.ijhydene.2025.151594}}.

\bibitem{19_wang2021hydrogenstorage}
Z.~Wang, Y.~Wang, S.~Afshan, J.~Hjalmarsson, A review of metallic tanks for
  {$\mathrm{H_2}$} storage with a view to application in future green shipping,
  International Journal of Hydrogen Energy 46~(9) (2021) 6151--6179.
\newblock \href
  {https://doi.org/https://doi.org/10.1016/j.ijhydene.2020.11.168}
  {\path{doi:https://doi.org/10.1016/j.ijhydene.2020.11.168}}.

\bibitem{20_afshan2023highperformance}
S.~Afshan, W.~Li, Z.~Wang, W.~Bailey, Y.~Wang, High-performance metallic
  materials for applications in infrastructure and energy sectors, Steel
  Construction 16~(3) (2023) 144--150.
\newblock \href {https://doi.org/https://doi.org/10.1002/stco.202300025}
  {\path{doi:https://doi.org/10.1002/stco.202300025}}.

\bibitem{IMO2015IGF}
{International Maritime Organization}, International Code of Safety for Ships
  Using Gases or Other Low-Flashpoint Fuels ({IGF} Code), IMO, London, 2015.

\bibitem{IMO2016IGC}
{International Maritime Organization}, International Code for the Construction
  and Equipment of Ships Carrying Liquefied Gases in Bulk (IGC Code), IMO,
  London, 2016.

\bibitem{chen2025hydrogen}
Y.-S. Chen, C.~Huang, P.-Y. Liu, H.-W. Yen, R.~Niu, P.~Burr, K.~L. Moore,
  E.~Mart{\'\i}nez-Pa{\~n}eda, A.~Atrens, J.~M. Cairney, Hydrogen trapping and
  embrittlement in metals – {A} review, International Journal of Hydrogen
  Energy 136 (2025) 789--821.
\newblock \href
  {https://doi.org/https://doi.org/10.1016/j.ijhydene.2024.04.076}
  {\path{doi:https://doi.org/10.1016/j.ijhydene.2024.04.076}}.

\bibitem{anoop2021review}
C.~R. Anoop, R.~K. Singh, R.~R. Kumar, M.~Jayalakshmi, T.~A. Prabhu, K.~T.
  Tharian, S.~V. S.~N. Murty, A review on steels for cryogenic applications,
  Materials Performance and Characterization 10~(2) (2021) 16--88.
\newblock \href {https://doi.org/https://doi.org/10.1520/MPC20200193}
  {\path{doi:https://doi.org/10.1520/MPC20200193}}.

\bibitem{sanmachi2008}
C.~San~Marchi, B.~P. Somerday, X.~Tang, G.~H. Schiroky, Effects of alloy
  composition and strain hardening on tensile fracture of hydrogen-precharged
  type 316 stainless steels, International Journal of Hydrogen Energy 33~(2)
  (2008) 889--904.
\newblock \href
  {https://doi.org/https://doi.org/10.1016/j.ijhydene.2007.10.046}
  {\path{doi:https://doi.org/10.1016/j.ijhydene.2007.10.046}}.

\bibitem{michler2008}
T.~Michler, A.~A. Yukhimchuk, J.~Naumann, Hydrogen environment embrittlement
  testing at low temperatures and high pressures, Corrosion Science 50~(12)
  (2008) 3519--3526.
\newblock \href {https://doi.org/https://doi.org/10.1016/j.corsci.2008.09.025}
  {\path{doi:https://doi.org/10.1016/j.corsci.2008.09.025}}.

\bibitem{takaki2016}
S.~Takaki, S.~Nanba, K.~Imakawa, A.~Macadre, J.~Yamabe, H.~Matsunaga,
  S.~Matsuoka, Determination of hydrogen compatibility for solution-treated
  austenitic stainless steels based on a newly proposed nickel-equivalent
  equation, International Journal of Hydrogen Energy 41~(33) (2016)
  15095--15100.
\newblock \href
  {https://doi.org/https://doi.org/10.1016/j.ijhydene.2016.06.193}
  {\path{doi:https://doi.org/10.1016/j.ijhydene.2016.06.193}}.

\bibitem{sanmachi2021}
C.~San~Marchi, J.~A. Ronevich, J.~E.~C. Sabisch, J.~D. Sugar, D.~L. Medlin,
  B.~P. Somerday, Effect of microstructural and environmental variables on
  ductility of austenitic stainless steels, International Journal of Hydrogen
  Energy 46~(23) (2021) 12338--12347.
\newblock \href
  {https://doi.org/https://doi.org/10.1016/j.ijhydene.2020.09.069}
  {\path{doi:https://doi.org/10.1016/j.ijhydene.2020.09.069}}.

\bibitem{Fukuyama2003}
S.~Fukuyama, D.~Sun, L.~Zhang, M.~Wen, K.~Yokogawa, Effect of temperature on
  hydrogen environment embrittlement of type 316 series austenitic stainless
  steels at low temperatures, Journal of the Japan Institute of Metals 67~(9)
  (2003) 456--459.
\newblock \href
  {https://doi.org/https://doi.org/10.1016/j.ijhydene.2008.02.021}
  {\path{doi:https://doi.org/10.1016/j.ijhydene.2008.02.021}}.

\bibitem{KOMATSU2021}
A.~Komatsu, M.~Fujinami, M.~Hatano, K.~Matsumoto, M.~Sugeoi, L.~Chiari,
  Straining-temperature dependence of vacancy behavior in hydrogen-charged
  austenitic stainless steel {316L}, International Journal of Hydrogen Energy
  46~(9) (2021) 6960--6969.
\newblock \href
  {https://doi.org/https://doi.org/10.1016/j.ijhydene.2020.11.148}
  {\path{doi:https://doi.org/10.1016/j.ijhydene.2020.11.148}}.

\bibitem{8_alvarez2023hydrogen}
G.~Álvarez, Z.~Harris, K.~Wada, C.~Rodríguez, E.~Martínez-Pañeda, Hydrogen
  embrittlement susceptibility of additively manufactured {316L} stainless
  steel: Influence of post-processing, printing direction, temperature and
  pre-straining, Additive Manufacturing 78 (2023) 103834.
\newblock \href {https://doi.org/https://doi.org/10.1016/j.addma.2023.103834}
  {\path{doi:https://doi.org/10.1016/j.addma.2023.103834}}.

\bibitem{Bak2017}
S.~H. Bak, S.~S. Kim, D.~B. Lee, Effect of hydrogen on dislocation structure
  and strain-induced martensite transformation in {316L} stainless steel, RSC
  Advances 7~(45) (2017) 27840--27845.
\newblock \href {https://doi.org/https://doi.org/10.1039/C7RA01053B}
  {\path{doi:https://doi.org/10.1039/C7RA01053B}}.

\bibitem{ISHTIAQ2025}
M.~Ishtiaq, Y.-K. Kim, S.~Tiwari, C.~H. Lee, W.~H. Jo, H.~Sung, K.-S. Cho,
  S.-G. Kang, Y.-S. Na, J.~B. Seol, Serration-induced plasticity in phase
  transformative stainless steel {316L} upon ultracold deformation at 4.2~k,
  Materials Science and Engineering: A 921 (2025) 147591.
\newblock \href {https://doi.org/https://doi.org/10.1016/j.msea.2024.147591}
  {\path{doi:https://doi.org/10.1016/j.msea.2024.147591}}.

\bibitem{Li2023}
S.~Li, P.~J. Withers, S.~Kabra, K.~Yan, The behaviour and deformation
  mechanisms for 316{L} stainless steel deformed at cryogenic temperatures,
  Materials Science and Engineering: A 880 (2023) 145279.
\newblock \href {https://doi.org/https://doi.org/10.1016/j.msea.2023.145279}
  {\path{doi:https://doi.org/10.1016/j.msea.2023.145279}}.

\bibitem{KIM2024}
M.~S. Kim, T.~Lee, Y.~Kim, Comparative analysis of unloading compliance and
  normalization methods for fracture toughness assessment in {316L} stainless
  steel at cryogenic temperatures, Theoretical and Applied Fracture Mechanics
  132 (2024) 104474.
\newblock \href {https://doi.org/https://doi.org/10.1016/j.tafmec.2024.104474}
  {\path{doi:https://doi.org/10.1016/j.tafmec.2024.104474}}.

\bibitem{HATANO2014}
M.~Hatano, M.~Fujinami, K.~Arai, H.~Fujii, M.~Nagumo, Hydrogen embrittlement of
  austenitic stainless steels revealed by deformation microstructures and
  strain-induced creation of vacancies, Acta Materialia 67 (2014) 342--353.
\newblock \href {https://doi.org/https://doi.org/10.1016/j.actamat.2013.12.039}
  {\path{doi:https://doi.org/10.1016/j.actamat.2013.12.039}}.

\bibitem{YAMABE2017}
J.~Yamabe, O.~Takakuwa, H.~Matsunaga, H.~Itoga, S.~Matsuoka, Hydrogen
  diffusivity and tensile-ductility loss of solution-treated austenitic
  stainless steels with external and internal hydrogen, International Journal
  of Hydrogen Energy 42~(18) (2017) 13289--13299.
\newblock \href
  {https://doi.org/https://doi.org/10.1016/j.ijhydene.2017.04.055}
  {\path{doi:https://doi.org/10.1016/j.ijhydene.2017.04.055}}.

\bibitem{4_bseniso6892-1}
{British Standards Institution}, {BS EN ISO 6892-1: Metallic materials -
  Tensile testing - Part 1: Method of test at room temperature} (2019).

\bibitem{BRASS2006}
A.~M. Brass, J.~Ch{\^e}ne, Hydrogen uptake in {316L} stainless steel:
  Consequences on the tensile properties, Corrosion Science 48~(10) (2006)
  3222--3242.
\newblock \href {https://doi.org/https://doi.org/10.1016/j.corsci.2005.11.004}
  {\path{doi:https://doi.org/10.1016/j.corsci.2005.11.004}}.

\bibitem{QUAN2026}
S.~Quan, A.~Zafra, E.~Martínez-Pañeda, C.~Wu, Z.~D. Harris,
  L.~Cupertino-Malheiros, New insights into hydrogen-assisted intergranular
  cracking in nickel, Materials Science and Engineering: A 950 (2026) 149545.
\newblock \href {https://doi.org/https://doi.org/10.1016/j.msea.2025.149545}
  {\path{doi:https://doi.org/10.1016/j.msea.2025.149545}}.

\bibitem{SantosMaldonado2024}
C.-T. Santos~Maldonado, A.~Zafra, E.~Martínez-Pañeda, P.~Sandmann, R.~Morana,
  M.-S. Pham, Influence of dislocation cells on hydrogen embrittlement in
  wrought and additively manufactured inconel 718, Communications Materials
  5~(1) (2024) 223.
\newblock \href {https://doi.org/https://doi.org/10.21203/rs.3.rs-4217438/v1}
  {\path{doi:https://doi.org/10.21203/rs.3.rs-4217438/v1}}.

\bibitem{ZAFRA2023_TDS}
A.~Zafra, Z.~Harris, E.~Korec, E.~Mart{\'i}nez-Pa{\~n}eda, On the relative
  efficacy of electropermeation and isothermal desorption approaches for
  measuring hydrogen diffusivity, International Journal of Hydrogen Energy
  48~(3) (2023) 1218--1233.
\newblock \href
  {https://doi.org/https://doi.org/10.1016/j.ijhydene.2022.10.025}
  {\path{doi:https://doi.org/10.1016/j.ijhydene.2022.10.025}}.

\bibitem{7_bseniso6892-4}
{British Standards Institution}, {BS EN ISO 6892-4: Metallic materials -
  Tensile testing: Part 4: Method of testing at low temperature} (2015).

\bibitem{diaz2025comsol}
A.~D{\'i}az, J.~M. Alegre, I.~I. Cuesta, E.~Mart{\'i}nez-Pa{\~n}eda, A {COMSOL}
  framework for predicting hydrogen embrittlement, {Part I}: Coupled hydrogen
  transport, Engineering Fracture Mechanics 319 (2025) 111007.
\newblock \href
  {https://doi.org/https://doi.org/10.1016/j.engfracmech.2025.111007}
  {\path{doi:https://doi.org/10.1016/j.engfracmech.2025.111007}}.

\bibitem{SanMarchi2012}
C.~San~Marchi, B.~P. Somerday, Technical reference on hydrogen compatibility of
  materials, Tech. rep., Sandia National Laboratories, Livermore, CA, matls
  Tech Ref, MS-9402, Sandia National Laboratories (2012).

\bibitem{10_bauchau2008structural}
O.~Bauchau, J.~Craig, Structural Analysis, 1st Edition, Springer, Dordrecht,
  2008.

\bibitem{11_zheng2018effect}
C.~Zheng, W.~Yu, Effect of low-temperature on mechanical behavior for an {AISI}
  304 austenitic stainless steel, Materials Science and Engineering: A 710
  (2018) 359--365.
\newblock \href {https://doi.org/https://doi.org/10.1016/j.msea.2017.11.003}
  {\path{doi:https://doi.org/10.1016/j.msea.2017.11.003}}.

\bibitem{Zheng2022low}
C.~Zheng, L.~Hu, Q.~Zhen, Y.~Tang, Y.~Wang, N.~Li, H.~Jiang, Low temperature
  mechanical behavior of fine- and ultrafine-grained 304 austenitic stainless
  steel fabricated by cryogenic-rolling and annealing, Materials
  Characterization 191 (2022) 112084.
\newblock \href {https://doi.org/https://doi.org/10.1016/j.matchar.2022.112084}
  {\path{doi:https://doi.org/10.1016/j.matchar.2022.112084}}.

\bibitem{16_kim2023tensile}
M.~Kim, T.~Lee, J.~Park, Y.~Kim, Tensile and fracture characteristics of {304L}
  stainless steel at cryogenic temperatures for liquid hydrogen service, Metals
  13~(10) (2023) 1774.
\newblock \href {https://doi.org/https://doi.org/10.3390/met13101774}
  {\path{doi:https://doi.org/10.3390/met13101774}}.

\bibitem{17_fernandezpison2021flow}
P.~Fernández-Pisón, J.~A. Rodríguez-Martínez, E.~García-Tabarés,
  I.~Avilés-Santillana, S.~Sgobba, Flow and fracture of austenitic stainless
  steels at cryogenic temperatures, Engineering Fracture Mechanics 258 (2021)
  108042.
\newblock \href
  {https://doi.org/https://doi.org/10.1016/j.engfracmech.2021.108042}
  {\path{doi:https://doi.org/10.1016/j.engfracmech.2021.108042}}.

\bibitem{WADA2025}
K.~Wada, M.~Komatsu, Y.~Ono, J.~Yamabe, H.~Enoki, T.~Iijima, Hydrogen-induced
  degradation of sus304 austenitic stainless steel at cryogenic temperatures,
  Materials Science and Engineering: A 927 (2025) 147988.
\newblock \href {https://doi.org/https://doi.org/10.1016/j.msea.2025.147988}
  {\path{doi:https://doi.org/10.1016/j.msea.2025.147988}}.

\bibitem{Yang2023}
H.~Yang, T.~T. Nguyen, J.~Park, H.~M. Heo, J.~Lee, U.~B. Baek, Y.-K. Lee,
  Temperature dependency of hydrogen embrittlement in thermally h-precharged
  sts 304 austenitic stainless steel, Metals and Materials International 29
  (2023) 303--314.
\newblock \href {https://doi.org/https://doi.org/10.1007/s12540-022-01232-6}
  {\path{doi:https://doi.org/10.1007/s12540-022-01232-6}}.

\bibitem{Niessen2021}
F.~Niessen, A.~A. Gazder, J.~Hald, M.~A.~J. Somers, Multiscale in-situ studies
  of strain-induced martensite formation in inter-critically annealed
  extra-low-carbon martensitic stainless steel, Acta Materialia 220 (2021)
  117339.
\newblock \href {https://doi.org/https://doi.org/10.1016/j.actamat.2021.117339}
  {\path{doi:https://doi.org/10.1016/j.actamat.2021.117339}}.

\bibitem{Obst1991}
B.~Obst, A.~Nyilas, Experimental evidence on the dislocation mechanism of
  serrated yielding in f.c.c. metals and alloys at low temperatures, Materials
  Science and Engineering: A 137 (1991) 141--150.
\newblock \href {https://doi.org/https://doi.org/10.1016/0921-5093(91)90328-K}
  {\path{doi:https://doi.org/10.1016/0921-5093(91)90328-K}}.

\bibitem{Tabin2019}
J.~Tabin, B.~Skoczen, J.~Bielski, Discontinuous plastic flow coupled with
  strain induced fcc--bcc phase transformation at extremely low temperatures,
  Mechanics of Materials 129 (2019) 23--40.
\newblock \href {https://doi.org/https://doi.org/10.1016/j.mechmat.2018.10.007}
  {\path{doi:https://doi.org/10.1016/j.mechmat.2018.10.007}}.

\bibitem{HARRIS2018}
Z.~D. Harris, S.~K. Lawrence, D.~L. Medlin, G.~Guetard, J.~T. Burns, B.~P.
  Somerday, Elucidating the contribution of mobile hydrogen-deformation
  interactions to hydrogen-induced intergranular cracking in polycrystalline
  nickel, Acta Materialiaialia 158 (2018) 180--192.
\newblock \href {https://doi.org/https://doi.org/10.1016/j.actamat.2018.07.043}
  {\path{doi:https://doi.org/10.1016/j.actamat.2018.07.043}}.

\bibitem{deMeloFreire2025}
R.~M. de~Melo~Freire, M.~Kimura, T.~Kawabata, Theoretical model for hydrogen
  environment embrittlement in metastable austenitic stainless steel, Materials
  \& Design 256 (2025) 114268.
\newblock \href {https://doi.org/https://doi.org/10.1016/j.matdes.2025.114268}
  {\path{doi:https://doi.org/10.1016/j.matdes.2025.114268}}.

\bibitem{KIRCHHEIM2004}
R.~Kirchheim, Solid solutions of hydrogen in complex materials, Vol.~59 of
  Solid State Physics, Academic Press, 2004, pp. 203--291.
\newblock \href {https://doi.org/https://doi.org/10.1016/S0081-1947(04)80004-3}
  {\path{doi:https://doi.org/10.1016/S0081-1947(04)80004-3}}.

\bibitem{13_ding2019modified}
H.~Ding, Y.~Wu, Q.~Lu, Y.~Wang, J.~Zheng, P.~Xu, A modified stress-strain
  relation for austenitic stainless steels at cryogenic temperatures,
  Cryogenics 101 (2019) 89--100.
\newblock \href
  {https://doi.org/https://doi.org/10.1016/j.cryogenics.2019.06.003}
  {\path{doi:https://doi.org/10.1016/j.cryogenics.2019.06.003}}.

\bibitem{14_ding2018tensile}
H.~Ding, Y.~Wu, Q.~Lu, P.~Xu, J.~Zheng, L.~Wei, Tensile properties and impact
  toughness of s30408 stainless steel and its welded joints at cryogenic
  temperatures, Cryogenics 92 (2018) 50--59.
\newblock \href
  {https://doi.org/https://doi.org/10.1016/j.cryogenics.2018.04.002}
  {\path{doi:https://doi.org/10.1016/j.cryogenics.2018.04.002}}.

\bibitem{ROZENAK1990}
P.~Rozenak, I.~M. Robertson, H.~K. Birnbaum, {HVEM} studies of the effects of
  hydrogen on the deformation and fracture of aisi type 316 austenitic
  stainless steel, Acta Metallurgica et Materialia 38~(11) (1990) 2031--2040.
\newblock \href {https://doi.org/https://doi.org/10.1016/0956-7151(90)90070-W}
  {\path{doi:https://doi.org/10.1016/0956-7151(90)90070-W}}.

\end{thebibliography}

%% else use the following coding to input the bibitems directly in the
%% TeX file.

%% \bibitem[Author(year)]{label}
%% Text of bib
\end{document}